\documentclass[11pt,fleqn,a4paper]{article}
\pdfoutput=1 % if your are submitting a pdflatex (i.e. if you have
\usepackage[colorlinks=true, linkcolor=black!50!blue, urlcolor=blue, citecolor=blue, anchorcolor=blue]{hyperref}
\usepackage[font=small,labelfont=bf,margin=0mm,labelsep=period,tableposition=top]{caption}
\usepackage[a4paper,top=3cm,bottom=2.5cm,left=2.5cm,right=2.5cm,bindingoffset=0mm]{geometry}
\usepackage[T1]{fontenc} % if needed
\usepackage[latin1]{inputenc}
\usepackage[english]{babel}
\usepackage[symbol]{footmisc}
\usepackage{pifont}
\pdfoutput=1
\usepackage{graphicx,mathtools}
\usepackage{epsfig,cite}
\usepackage{amssymb}
\usepackage{amsmath}
\usepackage{color}
\usepackage{amstext,alltt,setspace}
\usepackage{amsbsy}
\usepackage{array}
\usepackage{booktabs, makecell}
\usepackage{hyperref}
\usepackage{slashed}
\usepackage{url}
\usepackage{geometry}
\usepackage{amssymb,amsthm,cancel,hyperref,graphicx,xcolor}
\usepackage{picinpar,graphicx,xy}
\usepackage[subnum]{cases}
\usepackage{booktabs}
\usepackage{mathrsfs}
\usepackage{amsfonts}
\usepackage{latexsym}
\usepackage{booktabs,graphicx,hyperref,epsfig,multirow}
\usepackage{bm}
\usepackage{comment}
\usepackage{ulem}
\newcommand{\ks}{k\hspace{-1.2ex}/}

\def\bea{\begin{eqnarray}}
\def\eea{\end{eqnarray}}
\def\beq{\begin{equation}}  
\def\eeq{\end{equation}}

\def\as{\alpha_s}

\def\bea{\begin{eqnarray}}
\def\eea{\end{eqnarray}}
\def\beq{\begin{equation}}
\def\eeq{\end{equation}}
\def\ba{\begin{eqnarray}}
\def\ea{\end{eqnarray}}
\def\be{\begin{equation}}
\def\ee{\end{equation}}
\definecolor{darkgreen}{HTML}{008000}

\renewcommand{\Re}{\mathrm{Re}}

\def\({\left(}
\def\){\right)}
\def\[{\left[}
\def\]{\right]}
\def    \hepph  #1 {{\tt hep-ph/#1}}
\def    \hepex  #1 {{\tt hep-ex/#1}}
\long\def\symbolfootnote[#1]#2{\begingroup%
\def\thefootnote{\fnsymbol{footnote}}\footnote[#1]{#2}\endgroup}

\def\lapprox{\lower .7ex\hbox{$\;\stackrel{\textstyle <}{\sim}\;$}}
\def\gapprox{\lower .7ex\hbox{$\;\stackrel{\textstyle >}{\sim}\;$}}

\renewcommand{\(}{\left(}
\renewcommand{\)}{\right)}
\renewcommand{\Re}{\mathrm{Re}\:}

\graphicspath{{Images/}}
\begin{document}
\begin{flushleft}
%%%%%%%%%%%%%%%%%%%%%%%%%%%%%%%%%%%%%%%%%%%%%%%%
\begin{figure}[h]
\includegraphics[width=.2\textwidth]{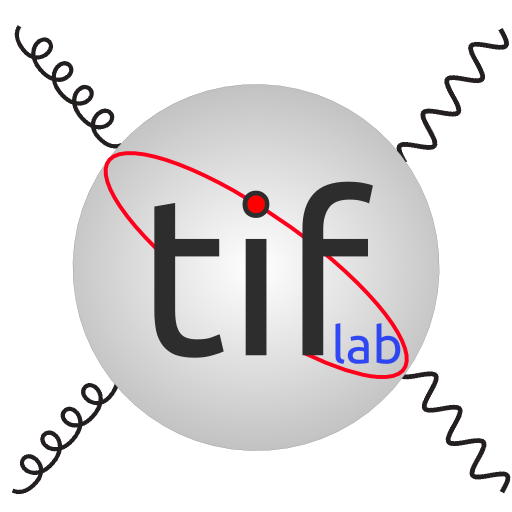}
\end{figure}
%%%%%%%%%%%%%%%%%%%%%%%%%%%%%%%%%%%%%%%%%%%%%%%%%
\end{flushleft}
\vspace{-5.0cm}
\begin{flushright}
TIF-UNIMI-2025-23
\end{flushright}

\vspace{2.0cm}

\begin{center}
{\Large \bf A simple introduction to soft resummation}
\end{center}

\vspace{1.3cm}

\begin{center}
Stefano Forte$^1$\footnote[1]{\it Speaker.} and Giovanni Ridolfi$^2$ \\
\vspace{.3cm}
{\it
{}$^1$Tif Lab, Dipartimento di Fisica, Universit\`a di Milano and\\
INFN, Sezione di Milano, Via Celoria 16, I-20133 Milano, Italy\\
{}$^2$Dipartimento di Fisica, Universit\`a di Genova and\\
INFN, Sezione di Genova, Via Dodecaneso 33, I-16146 Genova, Italy.\\
}
\vspace{1.3cm}

{\bf \large Abstract}
\end{center}
We provide an elementary pedagogical introduction to some basic concepts and
techniques of soft (or Sudakov) resummation, specifically in QCD,
paying particular attention to simple but useful tricks of the trade. We
briefly review collinear (Altarelli-Parisi) and infrared (eikonal)
factorization, cancellation of infrared singularities and
factorization of mass singularities. We recall basic concepts on
renormalization group invariance and the solution of renormalization
group equations. We then show how threshold resummation can be derived
from a renormalization group argument following from the cancellation
of infrared singularities. We discuss various equivalent forms of the
resummed result, and we briefly present transverse momentum
resummation.

\vspace{1.3cm}

\begin{center}
  {\it Lectures at the LXV Cracow School of Theoretical Physics\\
    Zakopane, Poland, June 2025}
  \end{center}

\clearpage
\tableofcontents
\clearpage
\section{Soft resummation from QED to QCD}
\label{sec:introduction}
The all-order resummation of logarithmically enhanced contributions
due to the emission of gauge vector bosons,
and specifically its use in the
context of QCD, is both a standard topic and an active research
field. The underpinnings of these
techniques in classic QED results are covered both in
old~\cite{Berestetskii:1982qgu} and more
recent~\cite{Sterman:1993hfp} graduate-level textbooks. Some recent
developments, such as for example webs~\cite{White:2015wha} have been
reviewed, and the approach to resummation based on soft-collinear
effective theories has been the subject of textbook
treatments~\cite{Becher:2014oda}. However, a treatment at the
introductory level of
the direct QCD approach, rooted in the original papers of Sterman~\cite{Sterman:1986aj},
Catani and Trentadue~\cite{Catani:1989ne} is rather more difficult to
find. It is the purpose of these elementary lectures to provide such
an introduction.

In order to make our treatment accessible and self-contained, we start
by reviewing textbook~\cite{Peskin:1995ev} material on the
factorization of soft and collinear emission, discussing the
cancellation of infrared singularities and the factorization of
collinear singularities. We then proceed to show the origin of
double-logarithmically enhanced contributions, and their all-order
exponentiation.  We take particular care in explaining some
simple computational tricks that are used in order to obtain the
desired results, for instance in the treatment of phase space. We then
review basic ideas on renormalization group 
invariance, the Callan-Symanzik renormalization group equation and its
solution, again at a basic textbook level. We then show how the
standard threshold resummation~\cite{Sterman:1986aj,Catani:1989ne} can
be derived using a renormalization-group
argument~\cite{Contopanagos:1996nh,Forte:2002ni}, explain the
underlying physical argument, illustrate the predictive power of the
resummed results, and demonstrate the equivalence of
various forms that it can take.  We conclude by briefly touching upon
transverse momentum resummation.
\clearpage

\begin{center}
  {\Large\bf Part I: soft and collinear logarithms}
\end{center}
\bigskip
Soft resummation in QCD sums to all perturbative orders contributions
that are enhanced by powers of the logarithm of the ratio of a soft
scale and the hard scale of the process. They arise because
contributions coming from real gluon emission diverge in the threshold
limit in which the soft scale vanishes. The divergence is canceled  at
threshold 
by virtual corrections, in which the emitted gluon is
absent, and the log is the left-over of the cancellation away from
threshold. This log gets combined with a second log arising from the
region of
integration over the transverse momentum of the emitted gluon in which
the emitted and emitting parton momenta become collinear
(i.e. parallel).
Resummation is possible because of the factorized nature of the
gluon emission process when the gluon is soft (i.e.  its energy tends
to zero) or collinear (i.e. its emission angle with respect  to the
emitting parton tends  
to zero). Multiple emissions can thus be treated as a branching
process with a suitable factorization of phase space.  We discuss in turn the factorized nature of
soft and collinear emission; singularities in the soft limit and the
logs they leave behind; and the exponentiation of multiple emission.

\section{Factorized emission}
\label{sec:scf}
The emission of soft gluons and the
emission of collinear partons (quark or gluons) from either quarks or
gluons factorizes, in the sense that the process with an extra gluon
or parton in the final state can be written in terms of that without
it, times a universal emission kernel. The factorization happens at
the level of amplitudes for soft gluons, and at the level of squared
amplitudes for collinear partons --- factorization of collinear
emission at the amplitude level is rather more subtle and goes beyond
the scope of our discussion (see 
Ref.~\cite{Duhr:2025cye}). We summarize main results and refer to
textbooks~\cite{Peskin:1995ev,Ellis:1996mzs} and reference
works~\cite{Collins:2011zzd,Weinberg:1995mt} for a more detailed discussion.
\subsection{Collinear emission}
\label{sec:coll}

\begin{figure}[!t]
 \centering
\includegraphics[width=.4\linewidth]{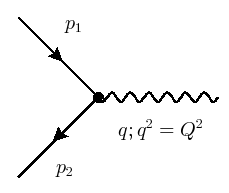}
\includegraphics[width=.4\linewidth]{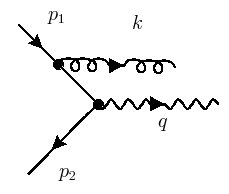}
\caption{The leading-order Drell-Yan production process (left) and a
  next-to-leading order real gluon emission correction to it (right).}    
 \label{fig:dy}
\end{figure}
%-------------------------------------------------------------------------------
We consider for definiteness the Drell-Yan process, i.e. production of
a gauge boson of mass (or virtuality) $Q^2$ in the annihilation of
a quark and antiquark with momenta $p_1$ and $p_2$ (see Fig.~\ref{fig:dy}). 
The leading-order amplitude for this process can be written
\begin{equation}
M_0(p_1,p_2)=\bar v(p_2)A_0(p_1,p_2)u(p_1).
\end{equation}
We consider a correction to the leading-order amplitude due to the
emission of an extra gluon with momentum $k$ from one of the two
incoming quark lines. It is convenient to introduce a Sudakov 
parametrization for the emitted gluon momentum. This consists of
writing it in terms of two light-like non-orthogonal vectors $p_1$ and
$p_2$, and a transverse vector $k_t$ orthogonal to both of them,
namely as
\begin{equation}\label{eq:ksud}
  k=(1-z)p_1+ k_t+\eta p_2,
\end{equation}
where
\begin{equation}\label{eq:basis}
   p_1^2=p_2^2=0;\quad p_1\cdot
  k_t=p_2\cdot
  k_t =0;\quad p_1\cdot p_2\not=0 .
\end{equation}

The value of $\eta$ is fixed by the on-shell condition $k^2=0$:
\begin{equation}\label{eq:etaval}
  \eta=\frac{-k_t^2/s}{1-z},
\end{equation}
where $s$ is 
the center-of-mass energy
\begin{equation}\label{eq:cms}
  (p_1+p_2)^2= 2 p_1\cdot p_2=s.
\end{equation}
The collinear limit we are interested in is the limit in which
$k_t^2\to 0$, so also $\eta\to0$.

As an example, for illustration,  take
\begin{align}\label{eq:sudex}
  p_1&=(p,0,0,p),\nonumber\\
  p_2&=(p,0,0,-p).
\end{align}
The value of $p$ is fixed by the center-of-mass energy
Eq.~(\ref{eq:cms}) to be given by
\begin{equation}\label{eq:cmsp}
 p=\frac{\sqrt s}{2}.
\end{equation}
Furthermore, 
\begin{equation}\label{eq:ktex}
  k_t=(0,\vec k_t,0),
\end{equation}
which shows that  $k_t^2\le 0$.

We assume for definiteness that the gluon is radiated by
the incoming quark, and we follow notation and conventions of
Ref.~\cite{Peskin:1995ev}, to which we also refer for the Feynman
rules of QCD.
We can write the corresponding contribution to the amplitude $M_1$ for  the process with the extra gluon
in terms of the tree-level amplitude $M_0$ as 
\begin{equation}\label{eq:m0tom1}
M_1^q=\bar v(p_2)A_0(p_1-k,p_2)\frac{i(\slashed p_1-\slashed k)}{(p_1-k)^2} i g_s\gamma^\mu\sum_a t^a\epsilon^a_\mu(k)u(p_1),
\end{equation}
where $t^a$ are SU(3) gauge group generators, and $\epsilon^a_\mu$ is the gluon polarization.
An analogous calculation yields the contribution of the emission from the antiquark line, $M_1^{\bar q}$.
We now note that in the collinear limit the momentum in
the intermediate propagator goes on shell. Indeed, we have
\begin{equation}\label{eq:intprop}
  (p_1-k)^2=- 2 p_1\cdot
  k=-s\eta=\frac{k_t^2}{1-z},
\end{equation}
where in the last step we have used Eq.~(\ref{eq:etaval}). But for an
on-shell (massless) fermion
\begin{equation}\label{eq:proj}
  \slashed p =\sum_s u^s(p)\bar u^s(p),
\end{equation}
where $ u^s(p)$ is the spin $s$ solution of the Dirac equation. 
Hence, expanding the intermediate fermion propagator about the
collinear limit, we have
\begin{equation}\label{eq:ww}
\frac{i(\slashed
    p_1-\slashed k)}{(p_1-k)^2}=i\frac{\sum_s u^s(p_1-k)\bar u^s(p_1-k)\left[1+O(k_t^2/s)\right]}{(p_1-k)^2},
\end{equation}
where it is understood that the argument $p_1-k$ of the spinor $u_s$
is an on-shell momentum, $(p_1-k)^2=0$, and the denominator is
expanded according to Eq.~(\ref{eq:intprop}) about the collinear
limit, in which it vanishes.

Substituting  Eq.~(\ref{eq:ww}) in the expression
Eq.~(\ref{eq:m0tom1}) of the amplitude 
and integrating over the
momentum of the radiated gluon we see that the unpolarized
squared matrix element, integrated over the phase space of the radiated gluon
\begin{equation}
d\Phi_k=\frac{d^3k}{(2\pi)^32E}
\end{equation} factorizes as 
\begin{equation}\label{eq:apfact}
\sum_s\int\left|M_1^q\right|^2d\Phi_k=\frac{\as}{2\pi}\int \frac{d|k_t^2|}{|k_t^2|} dz |M_0(p_1-k,p_2)|^2
P^r_{qq}(z)\left[1+O(k_t^2/s)\right],
\end{equation}
where we have defined $\alpha_s=\frac{g_s^2}{4\pi}$, $M_0(p_1-k,p_2)$ is
the tree-level Drell-Yan amplitude, but now with incoming quark
carrying momentum
\begin{equation}\label{eq:momscal}
  p_1-k= zp_1\left[1+O(k_t^2/s)\right],
    \end{equation}
and $P^r_{qq}(z)$ is a dimensionless function,
proportional to the squared matrix element for the tree-level emission
process $q(p_1)\to q(p_1-k)g(k)$.  The function  $P^r_{qq}(z)$ is
called a splitting function; the superscript $r$ denotes the fact that
this is only the contribution to the splitting function due to real
emission, while there is also a virtual contribution due to gluon
loops, that we will discuss in Sect.~\ref{sec:ircanc} below.
Note that the fact that the most
singular contribution in Eq.~(\ref{eq:apfact}) as $|k_t^2|\to0$ is
proportional to $\frac{1}{|k_t^2|}$, and the fact that  $P^r_{qq}$ is a
function of $z$ only are both entirely fixed by dimensional analysis,
as both $|M_0|^2$ and $\int |M_1^q|^2d\Phi_k$ have the same dimensions. The integral over $|k_t|$
is logarithmic; its upper limit is kinematically bounded by the center
of mass energy $s$, while at the lower limit it is divergent. The way
to deal with the divergence will be discussed later. 

Now, it should be clear that nowhere in the above argument the
explicit form of $M_0$ was used. So what we have actually proven is
that radiation of a gluon from an external quark line produces a
logarithmically enhanced contribution to the square amplitude, which is
proportional to the original amplitude, but with an incoming quark
momentum rescaled by a factor $z\le 1$, and a universal splitting
function, which
only depends on the radiation process. 
A similar argument also works for gluon radiation from an incoming
gluon line, because one may likewise rewrite in the collinear
limit the numerator of the intermediate gluon propagator as a sum over
on-shell gluon polarization vectors, up to subtleties related to the
choice of gauge in the nonabelian case. Of course, in this case the
splitting function will be  different:  $P^r_{gg}$ now depending on the
$g(p_1)\to g(p_1-k)g(k)$ matrix element.
Furthermore, what is being used in the argument is the on-shell
behavior of the
intermediate propagator. This means that the argument also works in
the off-diagonal case, except that in this case an incoming quark is
replaced by a gluon with factor $P^r_{gq}(z)$ and an incoming gluon is replaced by a
quark with factor $P^r_{qg}(z)$. Finally, it is clear that the argument
also works for massive quarks, though in this case the logarithmic
integral at the lower end is regulated by the quark mass.

In short, we have proven that the squared matrix element for radiation from an external
line produces a contribution which is enhanced by a logarithm of the
hard scale of the process, 
factorized in terms of universal splitting functions, up to terms with
a relative power suppression in the hard scale. The argument of the
log is the ratio of the hard scale to the radiator mass, and hence it
diverges in the  massless limit - a divergence which is usually  referred to
as collinear or mass singularity.

\begin{figure}[!t]
  \centering
  \begin{minipage}{.3\linewidth}\begin{center}\vglue2truecm
      \includegraphics[width=\linewidth]{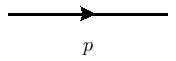}\end{center}\end{minipage}   \begin{minipage}{.2\linewidth}\begin{center}\vglue1.4truecm
      {\Huge $\Rightarrow$ } \end{center}\end{minipage} 
\begin{minipage}{.3\linewidth}\begin{center}
\includegraphics[width=\linewidth]{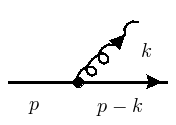}\end{center}\end{minipage} 
\caption{Radiation of a gluon with momentum $k$ from a quark with
  momentum $p$.}    
 \label{fig:eik}
\end{figure}
%-------------------------------------------------------------------------------
\subsection{Soft radiation}
\label{sec:soft}
Consider now the case in which a gluon with momentum $k$ is emitted
from an external  quark or antiquark with momentum $p$ and mass $m$,
and the energy of the gluon, and thus all the components of $k$, tend
to zero. In this case at the amplitude level, assuming for
definiteness that the emitting particle is an incoming quark,  the external wave
function factor $u(p)$ is supplemented by the insertion of a vertex
and a propagator factor (see Fig.~\ref{fig:eik}):
\begin{equation}\label{eq:softins}
  u(p)\quad\rightarrow\quad W_a^\mu(p,k) u(p)=  \frac{\slashed p+\slashed
    k+m}{(p-k)^2-m^2}g_s\gamma^\mu t^a u(p).
\end{equation}
%a$ are respectivel;y the Lorentz and color indices of
                %the emitted gluon.
Expanding about $k=0$ and using the Dirac equation
\begin{equation}\label{eq:softeik}
W_a^\mu(p,k) u(p)=-g_s\frac{2p^\mu-\gamma^\mu (\slashed
  p-m)}{2 p\cdot k}t^a
u(p)\left[1+O(k)\right]=-t^au(p)g_s \frac{p^\mu}{p\cdot
  k}\left[1+O(k)\right].
\end{equation}
The factor $\frac{p^\mu}{p\cdot
  k}$ is referred to as  eikonal factor, and this approximation as 
eikonal approximation.

Again, a similar argument works also for soft gluon emission from a
gluon line. However, unlike for
collinear emission, in the case of soft emission the argument only
works for diagonal emission, i.e. gluon emission from  a quark or gluon
line. On the other hand, now  factorization happens at the amplitude
level, and the eikonal factor is the same both for emission from a
quark and a gluon line. Also in this case, integration over the emitted gluon momentum leads
to a logarithm of the hard scale, divergent in the infrared as we now
discuss.

\section{Singularities and logs}
\label{sec:singlogs}

Integration over the momentum of the emitted gluon leads to a double
logarithm: a collinear log coming from the lower end of the integration over transverse
momentum, and an infrared log coming from the lower end of the
integration over energy. Both are divergent; however the way they
are treated is substantially different: the infrared divergence cancels against
a virtual contribution in the case of infrared logs, while 
collinear logs can be factored away exploiting their universality
properties.
\subsection{The Sudakov double log}
\label{sec:sudakov}
\begin{figure}[!t]
 \centering
\includegraphics[width=.3\linewidth]{zakoplots/dylo.png}
\includegraphics[width=.3\linewidth]{zakoplots/dynlob.png}
\includegraphics[width=.3\linewidth]{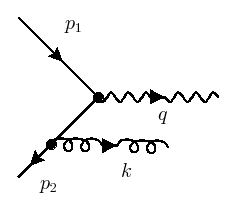}
\caption{The leading order Drell-Yan process and the first-order real-emission corrections to it.}    
 \label{fig:dyfull}
\end{figure}

In order to understand the origin of the double soft-collinear log,
consider again the full set of gluon emission corrections to the
tree-level Drell-Yan process (see Fig.~\ref{fig:dyfull}). We are
specifically interested in the form of the phase space integration
over the emitted gluon momentum $k$. To this purpose, it is
convenient to use a slightly different version of the Sudakov
parametrization, namely the light-cone parametrization,
in which instead of $p_1$ and $p_2$ we adopt as
basis vectors their linear combinations
\begin{equation}\label{eq:ppm}
  p^\pm=
  \frac{p_1\pm p_2}{2},
\end{equation}
namely
\begin{equation}\label{eq:ksud1}
k=(1-x) p^+ +y p^-+k_t.
\end{equation}

The on-shell condition now gives
\begin{equation}\label{eq:sudp}
  y=\pm\sqrt{(1-x)^2-\frac{4|k_t^2|}{s}},
\end{equation}
where the two solutions correspond to $k$ collinear to either $p_1$ or $p_2$ as $k_t^2\to0$.
The Lorentz invariant phase space integration measure
\begin{equation}\label{eq:phsp}
 d\Phi_k=\frac{|k_t|d |k_t|d\phi dk_z}{2 E (2\pi)^3}
\end{equation}
can be written in terms of $x$ and $k_t^2$ by rewriting the
longitudinal momentum integral as an energy integral through
\begin{equation}\label{eq:evskz}
  \frac{ dk_z}{E }=\frac{ dE}{|k_z|}
\end{equation}
and then noting that with the Sudakov parametrization
Eq.~(\ref{eq:sudp})
\begin{equation}\label{eq:kesud}
  k_z=y \frac{\sqrt s}{2}; \quad E=\frac{\sqrt s}{2}(1-x) .
\end{equation}
It follows that 
\begin{equation}\label{eq:phspsud}
d\Phi_k =\frac{1}{4 (4\pi^2)}\frac{ dx d |k_t^2|}{ \sqrt{(1-x)^2-\frac{4
      |k_t^2|}{s}}}.
\end{equation}

Because we are interested in the limit in which $k$ is both collinear
and soft, we can use the eikonal approximation
Eq.~(\ref{eq:softeik}). We include all first-order real emission
corrections, i.e. those corresponding both to radiation from the quark
and the antiquark, because the inclusion of the  full gauge-invariant set of
diagrams avoids some
subtleties that would instead arise when computing directly the
splitting function corresponding to the emission from a single quark or
antiquark line, as in Sect.~\ref{sec:coll}-\ref{sec:soft}. Of course,
the full set of diagrams also involves virtual loop corrections that to
this order interfere with the leading-order process: we will
come to these in Sect.~\ref{sec:ircanc}.

The sum of the two real emission corrections
of Fig.~\ref{fig:dyfull} to the amplitude is then
\begin{equation}\label{eq:nlodl}
M_1=M_1^q+M_1^{\bar q}= M_0 t^a g_s \left(\frac{p_2^\mu}{p_2\cdot k} -
  \frac{p_1^\mu}{p_1\cdot k}\right)\epsilon^*_\mu(k)\left[1+O(1-x)\right],
\end{equation}
which is correct (see Eq.~(\ref{eq:softeik})) up to terms of the order of
the gluon energy $E=\frac{\sqrt{s}}{2}(1-x)$,
and thus the square amplitude is
\begin{equation}\label{eq:nlosqua}
|M_1|^2=  |M_0|^2 C_F g^2_s
\frac{2p_1\cdot p_2}{(p_1\cdot k )\,(p_2\cdot k)}\left[1+O(1-x)\right],
\end{equation}
where the factor of $C_F$ comes from summing over all gluon colors and
using $\sum_a t^at^b=C_F \mathbb{I}$, and we have used the sum over
gluon polarizations
$\sum\epsilon^\mu{\epsilon^*}^\nu=-g^{\mu\nu}$, as it is
appropriate for a gauge-invariant amplitude.

Equation~(\ref{eq:nlosqua}) can be  written in terms of $x$ and $k_t$
using the Sudakov 
parametrization of $k$. We have
\begin{equation}\label{eq:sudscalp}
 p_1\cdot p_2=\frac{s}{2};\quad
 p_1\cdot k=  \frac{s}{4}[(1-x)-y];\quad p_2\cdot k=  \frac{s}{4}[(1-x)+y]
\end{equation}
so, adding up the two contributions corresponding to the two choices of
sign for $y$, Eq.~(\ref{eq:sudp}), we get
\begin{equation}\label{eq:nlosquf}
|M_1|^2= |M_0|^2C_F g_s^2\frac{32}{s[(1-x)^2-y^2]}\left[1+O(1-x)\right]=|M_0|^2  8C_F g_s^2
\frac{1}{|k_t^2|}\left[1+O(1-x)\right].
\end{equation}
The integration over the phase space of the radiated gluon yields
\begin{equation}\label{eq:dl1}
\int   d\Phi_k|M_1|^2= |M_0|^2 \frac{2\alpha_s}{\pi} \int \frac{ dx }{ \sqrt{(1-x)^2-
\frac{4|k_t^2|}{s}}} \frac{d |k_t^2|}{|k_t^2|}\left[1+O(1-x)\right].
\end{equation}
The origin of the double log is now clear: there is a logarithmic
integration over $k_t^2$, and furthermore  when
$k_t\to 0$ the energy denominator in the phase space reduces to
$\sqrt{(1-x)^2-\frac{4|k_t^2|}{s}}=(1-x)\left[1+O(|k_t^2|/s)\right]$, which
  leads to a further logarithmic integration over $x$, i.e. over
  energy  (recall Eq.~(\ref{eq:kesud})).
  
  The double integral can be treated using the distributional identity
  \begin{equation}\label{eq:distktx}
\frac{1}{ \sqrt{(1-x)^2-\frac{4
      |k_t^2|}{s}}}
=\left[\frac{1}{(1-x)_+}-\frac{1}{2}\delta(1-x)
\ln\frac{4|k_t^2|}{s}\right]\left[1+O(|k_t^2|/s)\right].
\end{equation}
Here the plus distribution is defined by its action on a generic test
function $f(x)$ upon integration over $x$: 
 \begin{equation}\label{eq:plus} 
\int_0^1 dx
\frac{1}{(1-x)_+} f(x)\equiv\int_0^1 dx
\frac{ f(x)-f(1)}{1-x}.
 \end{equation}
 Note that by construction
\begin{equation}\label{eq:plus0}
  \int_0^1dx\,\frac{1}{(1-x)_+}=0.
\end{equation}

Using Eq.~(\ref{eq:plus0}) it is apparent that only the second term in
Eq.~(\ref{eq:distktx}) contributes to the integral, which is then
straightforwardly giving a squared logarithm  (usually called double log). 
The upper limit of integration
is the maximum allowed value of $k_t^2$, which is proportional to the
center-of-mass energy $s$, while at the lower limit the integral
diverges, so we must introduce some infrared cutoff $\mu^2$ in order to
make sense of it. The treatment of this double divergence -- one
collinear and one soft -- will be the subject of the next two sections.
We thus get
\begin{equation}\label{eq:suddl}
\int   d\Phi_k|M_1|^2= C_F|M_0|^2 \frac{\alpha}{2\pi}
\ln^2\frac{s}{\mu^2}+  \hbox{less singular} \, ,
 \end{equation}
where we have neglected all contributions that are less singular as
$\mu^2\to0$. This includes both  the $O(1-x)$ corrections to the eikonal
approximation, which would in fact  be suppressed by inverse powers of
the dimensionful variable $s$, which contains the large scale of the
process, but also lower powers of log, which arise when expressing
the maximum value of $k_t^2$ in terms of $s$. For example, for the
simple $2\to2$ process of Fig.~\ref{fig:dy} one has ${k_t^2}_{\rm
  max}=s/4$ so $\ln^2 {k_t^2}_{\rm max}= \ln^2s- \ln4 \ln s+\text{non
  log}$. 

Note that again in the whole argument we never used the explicit form
of the zero-emission amplitude $M_0$, and also, that the eikonal
approximation also holds for emission of gluons from gluons. Hence we
conclude that in full generality the emission of
soft gluons from external lines of a generic amplitude leads to a
double soft and collinear log when integrating over the phase space of
the emitted gluon. In order to make sense of the result, however, we
must understand how to treat the soft and collinear singularity,
i.e. we have to get rid of the cutoff $\mu^2$.

\subsection{Mass singularities and their factorization}
\label{sec:massing}

As mentioned in Sect.~\ref{sec:coll}, collinear emission from a
massive parton factorizes with the same universal splitting functions,
but the collinear singularity is then regulated by the mass. Indeed,
in such  case one finds that the single-emission cross section is proportional to
\begin{align}\label{eq:masslog}
\int |M_1^q|^2 d\Phi_k
&=\frac{\as}{2\pi}\int_0^{{k_t^2}_{\rm max}}\frac{dk_t^2}{k_t^2-m^2} \int_0^1 dx P^r_{qq}(x)|M_0(xp_1,p_2)|^2
\nonumber\\
&=\frac{\as}{2\pi}\ln\frac{{k_t^2}_{\rm max}}{m^2} \int_0^1 dx P^r_{qq}(x)
  |M_0(xp_1,p_2)|^2+\hbox{ non log}.
\end{align}
The same conclusion holds if the radiating parton is off-shell,
because then the denominator of the intermediate propagator, 
Eq.~(\ref{eq:intprop}), contains a
contribution equal to the off-shellness $p_1^2$, and thus it does not
vanish when $k_t^2\to0$.

This observation is a key ingredient in solving the problem of the
collinear singularity. Indeed, quarks and gluons of course cannot
exist as free particles, and therefore an incoming parton is always
bound in a parent hadron, and consequently it is not a free massless
particle, but rather it has an off-shellness of the order of the
characteristic scale of the bound state. This however seems to trade a
problem -- the collinear singularity -- for a worse one -- the result
depends on a  scale which is determined by the nonperturbative physics
that describes the binding of quarks and gluons inside hadrons.

The way out emerges when combining the correction due to gluon
emission with the leading-order square 
matrix element $|M_0|^2$. Considering for simplicity the case of the
leading-order Drell-Yan process of Fig.~\ref{fig:dy}, the leading
order amplitude is fixed by momentum conservation. Defining
\begin{equation}\label{eq:taudef}
  \tau\equiv\frac{Q^2}{s}
\end{equation}
it  is clear that  the leading-order cross section $\sigma_0$ is proportional to $\delta(1-\tau)$ by
%$|M_0(p_1,p_2)|^2\propto\delta (1-\tau)$ by
energy-momentum conservation. After gluon emission, momentum $p_1$ and
thus the center-of-mass energy are rescaled by a factor $x$, hence
imposing $xs=Q^2$, the mass of the final-state gauge boson,
leads to $\tau=x$, so 
\begin{equation}\label{eq:nlox}
|M_0|^2\delta(1-\tau)+\int |M_1^q|^2d\Phi_k=
   \sigma_0\left[\delta (1-\tau)+ \frac{\as}{2\pi}P^r_{qq}(\tau)
   \ln\frac{{k_t^2}_{\rm max}}{m^2}\right]+\hbox{non log}.
\end{equation}
But now we note  that we can rewrite
Eq.~(\ref{eq:nlox}) as
\begin{align}\label{eq:massfact}
&|M_0|^2\delta(1-\tau)+\int|M_1^q|^2d\Phi_k
\nonumber\\
&=\sigma_0\int_\tau^1\frac{dx}{x} \left[\delta\left(1-\frac{\tau}{x}\right)+
  \frac{\as}{2\pi}P^r_{qq}\left(\frac{\tau}{x}\right)\ln\frac{{k_t^2}_{\rm max}}{\mu_F^2}\right]\left[\delta (1-x)+
  \frac{\as}{2\pi}P^r_{qq}(x) \ln\frac{\mu_F^2}{m^2}\right]
  \nonumber\\
  &+\hbox{non log}+O(\alpha_s^2),
\end{align}
where we introduced a factorization scale $\mu_F$, on which however
the result does not depend, up to
$O(\alpha_s^2)$ corrections.
We further observe that the hard  cross section that we have computed in
perturbation theory will always be combined with a parton distribution
function (PDF), which provides the appropriate weighting factor for
computing the cross-section with an incoming parton that carries a
momentum  $p_1$ related to the momentum  of a parent hadron $P_1$ as
$p_1=x_1P_1$.

We then note that 
the first factor  in square brackets in
Eq.~(\ref{eq:massfact}) is identical to our original
expression Eq.~(\ref{eq:nlox}), but now with
$\frac{\tau}{x}=1$, which corresponds to  $x s=Q^2$, as appropriate  for an
incoming quark with momentum $xp_1$, and also, with $m^2$ replaced by
$\mu_F^2$. Hence, we can
include the second factor in square brackets in Eq.~(\ref{eq:massfact})
in the PDF, effectively including in the PDF all radiation up to scale
$\mu_F^2$ that has downgraded the incoming quark momentum to $x$ times
what it was before radiation, and endowed the remaining parton with
virtuality $\mu_F^2$. 
We are thus left with a subtracted partonic cross section
\begin{equation}\label{eq:sigmahat}
  \hat\sigma_1^q(z) = \sigma_0\left[\delta (1-z)+
\frac{\as}{2\pi}P^r_{qq}\left(z\right) \ln\frac{{k_t^2}_{\rm max}}{\mu_F^2}+\hbox{non log}\right],
\end{equation}
where $z=\frac{Q^2}{xs}$ as it should be.

Because the splitting function is universal, this redefinition of the
PDF does not depend on the specific process, so the PDFs remain
universal -- a property of the hadron, and not of the process. On the
other hand, the 
partonic cross-section is now both finite and independent of
nonperturbative physics, but it has  acquired a dependence of the arbitrary
factorization scale $\mu_F$. However, this dependence is
compensated by an equal and
opposite dependence of the PDF, so the physical observables remain
independent of it.

\subsection{Cancellation of infrared singularities}
\label{sec:ircanc}
After subtraction of the collinear singularity -- which has been
factorized into the PDF, where it is then regulated by the
nonperturbative physics which determines the PDF -- we are still left
with an infrared singularity.
This can be seen clearly by looking at the explicit form of the
splitting function $P^r_{qq}$:
\begin{equation}\label{eq:pqq}
  P^r_{qq}(x)=C_F\frac{1+x^2}{1-x}.
\end{equation}
Recalling the form Eq.~(\ref{eq:masslog}) of the squared matrix
element, it is clear that as $x\to1$ the matrix element $|M_0(xp_1,p_2)|^2$ just reduces to its (regular) value that it had
without emission, so the integral over $x$ is logarithmically
divergent because of the denominator in Eq.~(\ref{eq:pqq}). Also
recalling that $x\to1$ is the limit in which the energy Eq.~(\ref{eq:kesud})
of the radiated gluon vanishes, this divergence is recognized as the
infrared singularity discussed in Sects.~\ref{sec:soft},\ref{sec:sudakov}.

\begin{figure}[!t]
  \centering
  \begin{minipage}{.24\linewidth}\begin{center}\vglue2truecm
      \includegraphics[width=\linewidth]{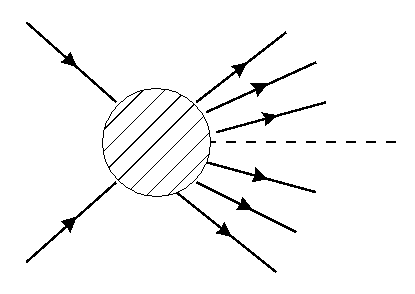}\\{\Large (a)}\end{center}\end{minipage}  \begin{minipage}{.24\linewidth}\begin{center}\vglue2truecm
      \includegraphics[width=\linewidth]{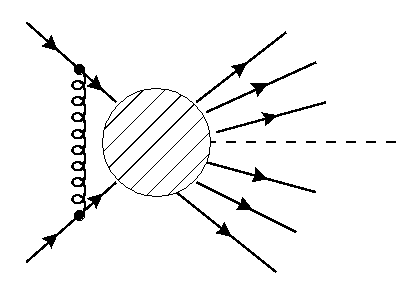}\\{\Large (b)}\end{center}\end{minipage}  \begin{minipage}{.24\linewidth}\begin{center}\vglue2truecm
      \includegraphics[width=\linewidth]{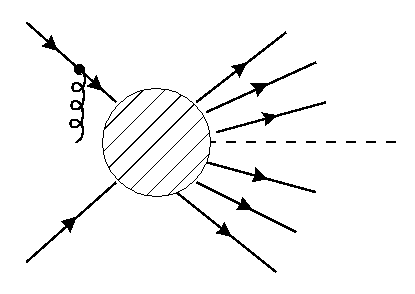}\\{\Large (c)}\end{center}\end{minipage}  \begin{minipage}{.24\linewidth}\begin{center}\vglue2truecm
      \includegraphics[width=\linewidth]{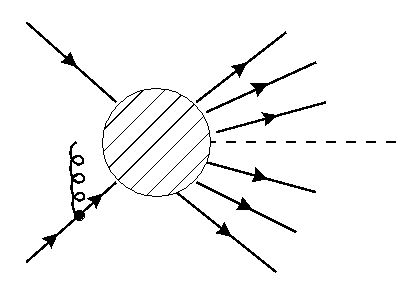}\\{\Large (d)}\end{center}\end{minipage}
\caption{Cancellation of infrared singularities: the interference
  between the leading-order process (a) and the loop correction (b)
  cancels the infrared singularity arising due to interference between
the two real-emission contributions (c) and (d).}
 \label{fig:kln}
\end{figure}

A classic result of quantum electrodynamics, the
Kinoshita-Lee-Nauenberg (KLN) theorem, guarantees that once one consistently
includes loop corrections along with real emission contributions] this
singularity cancels. The argument generalizes in QCD for colorless
final states~\cite{Weinberg:1995mt}.
The cancellation is illustrated graphically in
Fig.~\ref{fig:kln}.  The argument of
Sect.~\ref{sec:sudakov} shows that the infrared divergent contribution
to the cross-section can be obtained from the interference  of
the contributions to the amplitude coming from
gluons emitted from different quark lines lines.  Indeed,
the infrared singularity comes from the
$x\to1$ limit of the denominator in the amplitude
Eq.~(\ref{eq:nlosquf}), which in turn arises from the interference of
the two eikonal factors proportional to $p_1^\mu$ and $p_2^\mu$ in
Eq.~(\ref{eq:nlodl}). The KLN theorem ensures that this singularity
cancels against an equal and opposite singularity, which is present in
the virtual contribution that corresponds to the gluon loop obtained
by connecting the endpoints of the two interfering emitted real
gluons. This loop correction has of course the same final state
particles as the amplitude without either of the two interfering
final-state real emitted gluons, and thus it interferes with it when
calculating the square modulus of the amplitude, thereby producing a
contribution of the same order of the square of either of the real
emission terms (see Figure~\ref{fig:kln}).

A proof of the KLN theorem and its extension to QCD goes beyond the
scope of our discussion. Suffice it to say that it can be explicitly
verified, and indeed inclusion of both the real emission and the
loop correction leads to a splitting function which is free of
infrared singularities and takes the form
\begin{equation}\label{eq:pqqfull}
  P_{qq}(x)=C_F\left[\frac{1+x^2}{(1-x)_+}+\frac{3}{2}\delta(1-x)\right].
\end{equation}
Note that in terms of the single gluon emission process that is used
in order to compute the splitting function, the  loop correction has
the kinematics of the 
process without emission. Therefore, the contribution coming from  it
corresponds to the momentum of the parton before and after emission
being the same, i.e. it is localized at $x=1$: it only contributes to
distributions at $x=1$ such as the delta and the plus distribution in
Eq.~(\ref{eq:pqqfull}). 

The plus distribution in Eq.~(\ref{eq:pqqfull}) integrates to zero
according to Eq.~(\ref{eq:plus0}),  
but it leads to  logarithmically enhanced
contribution when the partonic cross-section is combined with a parton
distribution, as we shall see explicitly in Sect.~\ref{sec:softlimm}.
Indeed, the
virtual correction cancels the singularity at the endpoint $x=1$ but it
leaves behind a $\frac{1}{1-x}$ behavior away from the endpoint that
leads to logarithmic behavior upon integration over the momentum
fraction $x$. 

After factorization of the collinear singularity and cancellation of
the infrared singularity we are thus left with a finite
double-logarithmic contribution. Indeed, as we shall show explicitly
in Sect.~\ref{sec:softlim}, the upper limit of the transverse momentum
integral is
\begin{equation}\label{eq:ktmaxy}
{k_t^2}_{\rm max}=Q^2\frac{(1-z)^2}{4 z}.
\end{equation}
Substituting this in the expression Eq.~(\ref{eq:sigmahat}) of the
partonic cross-section, and adding the virtual contribution, one ends
up with 
\begin{equation}\label{eq:dlog}
\hat\sigma(z)=\sigma_0\left[\delta (1-z)+ \frac{\as}{2\pi} \left(P_{qq}(z)
\ln\frac{Q^2}{\mu_F^2}+\left[P^r_{qq}(z)\ln\frac{(1-z)^2}{4    z}\right]_+\right)\right].
\end{equation}
Hence, the partonic cross-section contains a double-logarithmic
contribution, coming from  interplay between the leftover of the infrared singularity,
that produces a  $\frac{1}{1-z}$ behavior, and the leftover of the
collinear singularity, that produces a $\ln{k_t^2}_{\rm  max}$
behavior, which in turns is sensitive to the infrared scale $Q^2(1-z)^2$.

\section{Multiple emission and exponentiation}
\label{sec:expon}
So far we have considered a single emission. Resummation sums multiple
emission to all orders, and it is thus useful to understand what happens
when several consecutive emissions take place. In such  case, it is necessary
to take into account the overall momentum conservation of the
radiation process. Once this is done, it turns out that multiple
emission leads to higher order powers of logs from a suitable region of
the final phase-space integration.

\subsection{Momentum conservation and phase space factorization}
\label{sec:phsp}
\begin{figure}[!t]
 \centering
\includegraphics[width=.7\linewidth]{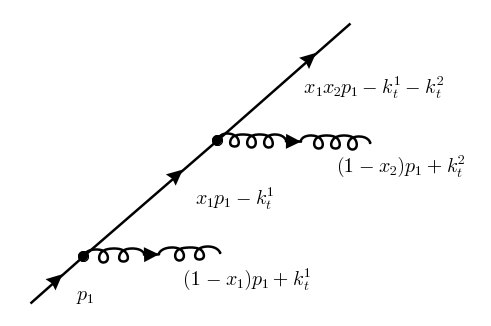}
\caption{Kinematics of multiple emission.}    
 \label{fig:multiple}
\end{figure}
In a multiple emission process (see Fig.~\ref{fig:multiple}) an incoming parton
with momentum $p_1$ emits a first gluon with momentum $k_1$ and the
outgoing parton after the first
emission has  momentum given by $p_1-k$. This  is then the incoming parton
for the second emission, and so forth. Using the Sudakov
parametrization Eq.~(\ref{eq:ksud}) it is clear that after the first
emission of a gluon with momentum
\begin{equation}\label{eq:ksudtwo}
  k_1=(1-x_1)p_1+ k_{t1}+\eta_1 p_2,
\end{equation}
the quark longitudinal momentum has become $x_1p_1$ and its transverse
momentum $- k_{t1}$. After the second emission the quark longitudinal momentum is
$x_1x_2 p_1$ and the transverse
momentum $- k_{t1}-k_{t2}$, and so on.

In a  total cross-section one integrates over the transverse momentum of
the final state, namely the transverse component of $q$,  the momentum of
the final state gauge boson, in the case of Drell-Yan production shown
in Fig.~\ref{fig:dy}. Therefore, all the integrals over transverse
momentum of the emitted gluons can be taken as independent, by simply
shifting the integration variable. On the other hand, the longitudinal
momentum is fixed by the condition that the final momentum $p_{n+1}$ after
$n$ emissions must take the value that is necessary in order to
produce the desired final state. So for instance in Drell-Yan
production after one collinear emission from the parton with momentum
$p_1$ the center-of-mass energy of the collision that produces the
gauge boson is (recall Eq.~(\ref{eq:taudef}) $2x_1 p_1\cdot
p_2=Q^2=s\tau$, after two emissions it is $2x_1 x_2 p_1\cdot
p_2=Q^2=s\tau$, and so on.

Therefore, when combining two subsequent emissions, the two splitting
functions that appear in the partonic cross-section according to
Eq.~(\ref{eq:sigmahat}), in order to guarantee longitudinal momentum
conservation,  must combine according to 
\begin{align}\label{eq:conv}
  \int dx_1dx_2\delta(x_2x_1-\tau)P_{qq}(x_1)P_{qq}(x_2)&=\int_0^1
           \frac{dx_1}{x_1}
           dx_2\delta\left(x_2-\frac{\tau}{x_1}\right)P_{qq}(x_1)P_{qq}(x_2)\nonumber\\
           &=\int_\tau^1
           \frac{dx_1}{x_1}
           P_{qq}\left(\frac{\tau}{x_1}\right)P_{qq}(x_1)\equiv
           \left[P_{qq}\otimes P_{qq}\right](\tau),
           \end{align}
where in the last step we have defined the convolution symbol $\otimes$.

It is clear that, consequently, the longitudinal momentum integrations
for the separate gluon emissions are not independent, because of the
momentum conservation constraint: phase space does not
factorize. However, it does factorize if one takes a suitable integral
transform: the Mellin transformation, that turns convolutions into
ordinary products. This means that if we define the anomalous
dimension
\begin{equation}\label{eq:andim}
\gamma_{qq}(N)\equiv \int_0^1 dx\,x^{N-1}
P_{qq}(x),
\end{equation}
then the convolution in Eq.~(\ref{eq:conv}) transforms into an ordinary product:
\begin{equation}\label{eq:convth}
\int_0^1 d\tau\,\tau^{N-1}\left[P_{qq}\otimes
P_{qq}\right](\tau)= \gamma_{qq}(N)\gamma_{qq}(N).
\end{equation}
This implies that by taking a Mellin transform we can treat emissions as
independent, which will then enable us to actually sum them to all orders.
\subsection{Ordered regions}
\label{sec:order}

We have seen that the transverse momentum integrations can be taken as
independent. However, we have also seen in Sect.~\ref{sec:massing} that if
the collinear radiation happens from an off-shell parton, then 
the off-shellness cuts off the momentum integration in the infrared,
thereby providing the lower scale of the collinear log. However, this
off-shellness is of course in turn given by transverse momentum of the
previous emission according to Eq.~(\ref{eq:intprop}). Hence the lower
limit of the transverse momentum integral for the second emission is
set by the value of the transverse momentum from the previous emission
and so on:
\begin{equation}\label{eq:double}
  \int_{\mu^2}^{{k_t^2}_{\rm max }}\frac{dk_{t1}^2}{k_{t1}^2}
  \int_{k_{t1}^2}^{{k_t^2}_{\rm max }}\frac{dk_{t2}^2}{k_{t2}^2}=
  \int_{\mu^2}^{{k_t^2}_{\rm max}}\frac{dk_{t1}^2}{k_{t1}^2}\ln\frac{{k_t^2}_{\rm max  }}{k_{t1}^2}=\frac{1}{2} \ln^2\frac{{k_t^2}_{\rm max}}{\mu^2}.
\end{equation}
Note that, of course, the integral over $k_{t2}^2$, the transverse
momentum of the second emission, also includes the region in which
$k_{t2}^2< k_{t1}^2$, but in this region there is no collinear
singularity because the singularity of the propagator after the second emission is
screened by the larger virtuality of the first emission, so the contribution from this integration  region
is not logarithmic.

The argument generalizes to the case of $n$ emissions:
\begin{equation}\label{eq:ntuple}
  \int_{\mu^2}^{{k_t^2}_{\rm max }}\frac{dk_{t1}^2}{k_{t1}^2}\dots
  \int_{k_{tn-1}^2}^{{k_t^2}_{\rm max }}\frac{dk_{tn}^2}{k_{tn}^2}=\frac{1}{n!} \ln^n\frac{{k_t^2}_{\rm max}}{\mu^2}.
\end{equation}
In Mellin space each subsequent emission is accompanied by a factor of
the anomalous dimension, and because of the factorization
Eq.~(\ref{eq:convth}) the $n$-emission term simply contains the
$n$-th power of the anomalous dimension
Hence the sequence of collinear emissions in the ordered transverse
momentum region in which  $k_{t1}^2<k_{t2}^2<\dots<k_{tn}^2$
exponentiates in Mellin space. Resummation to all orders of collinear
emission then leads to an exponential series of logarithmic
contributions.

The factorization  of the logarithms
and subtraction of the collinear
singularity according to Eq.~(\ref{eq:massfact}) then simply amounts
to splitting this exponentiated result into the product of two
exponentials, one including all transverse momentum integrations above
$\mu_F^2$, included in the partonic cross-section, and the other with
the region below $\mu_F^2$ included in the PDF. The dependence of the
PDF on the scale $\mu_F^2$ is the well-known Altarelli-Parisi 
evolution.

Soft resummation, that we now turn to, amounts to showing that the
double logs which are present in the soft limit, in which emitted
gluons are not only collinear but also soft, also exponentiate. The
exponentiation of collinear logs coming from
multiple collinear emission, that we have derived
here by studying the kinematics of the emission process, can be derived
using a renormalization group argument. A similar renormalization
group argument can then be used in order to also obtain the
exponentiation and resummation of infrared logs.

\clearpage

\begin{center}
  {\Large\bf Part II: resummation from renormalization group improvement}
\end{center}

The exponentiation of collinear logs that we discussed at the end of
the previous section can be understood as a consequence of the
independence of physical predictions on the factorization scale that
we introduced in Sect.~\ref{sec:massing}. The requirement of
independence can be cast in the form of a differential equation which
is in fact an instance of the general renormalization group equation
that expresses the scale dependence of predictions in quantum field theory. 
Soft resummation, i.e. the exponentiation of soft logs,
can be also derived from a renormalization group argument, which
generalizes the renormalization group argument leading to collinear
exponentiation.
%, to a case in which there are two
%classes of singularities with associate logarithms.

\section{Renormalization group  basics}
\label{sec:rgintro}
The basic idea of renormalization is that  in a
quantum field theory all couplings and fields must be defined in terms
of a reference scale, in such a way that all quantum fluctuations at
distances shorter than that of the reference scale are absorbed in the
definition of the couplings and fields. The fields and couplings must
consequently be defined by relating them to a reference scale. 
The quantum field theory then
provides relations between observable quantities, with the couplings
and fields used as a way to relate observables to each other. Infinities only
appear if one tries to express observable quantities in terms of
unobservable parameters defined at infinitely short distance scales.

For example, one may define the electron  charge  in quantum
electrodynamics in terms of the Coulomb force at a large distance,
i.e. at the scale of the (rest) electron mass. The (say) elastic
electron-electron scattering cross-section is finite once re-expressed
in terms of the electron charge thus defined. Renormalization group
invariance is then the  consequence of the fact that physical
predictions do not depend on the renormalization scale that has been
chosen to define the couplings. Because the dependence of the
couplings on this scale is a universal property of the theory, this
independence expresses the scale dependence of physical observables in
terms of this universal scale dependence of the coupling.

The basic idea of deriving  collinear exponentiation from
renormalization group invariance, that we will present shortly,
takes this one step further, based on
the observation that in QCD physically observable cross-sections are
expressed in factorized form in terms of parton distributions,
convoluted with partonic cross-sections. This factorization requires
introducing a scale, as discussed in Sect.~\ref{sec:massing} for the
subtraction of collinear singularities. As discussed in
Sect.~\ref{sec:order}, this can be viewed as a scale that splits in two the
parton radiation process, with radiation below the scale included in
the parton distribution, and radiation above it included in  the hard
matrix element.  Physical predictions are independent of this scale,
hence this scale dependence must cancel between the PDF and the hard
cross-section: indeed we have seen that it is expressed in terms of a
universal, process-independent splitting function, or rather, its
Mellin transform, the anomalous dimension Eq.~(\ref{eq:andim}).
But the PDF and the hard cross-section depend on different
variables: the hard cross-section depends on the specific hard scale
of the process, while the PDF does not. As we shall see,
dimensional analysis
then implies that the cancellation of the  dependence on the factorization scale can only
go through the scale dependence of the coupling constant: it is
calculable in terms of it, and the universal anomalous dimension. It can then
be integrated away, leading to all-order resummation and
exponentiation of logs of the hard physical scale.

\subsection{Renormalization group invariance and the renormalization
  group equation}
\label{sec:rge}
\begin{figure}[!t]
 \centering
\includegraphics[width=.24\linewidth]{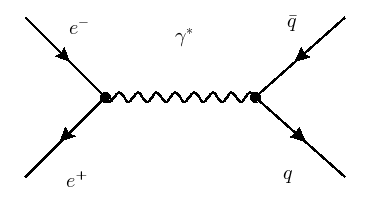}
\includegraphics[width=.24\linewidth]{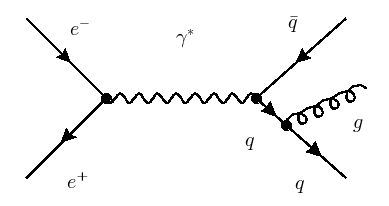}
\includegraphics[width=.24\linewidth]{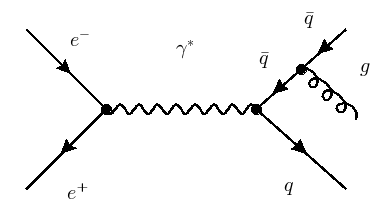}
\includegraphics[width=.24\linewidth]{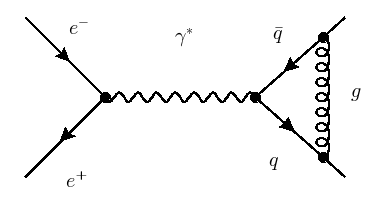}
\caption{Feynman diagrams contributing to the the production of
  partonic final states up to NLO in QCD.}    
 \label{fig:r}
\end{figure}
The simplest application of renormalization group invariance is 
to a physical observable that depends on a single scale. The
prototypical example is the $R$ ratio, defined as the 
cross-section for an electron-positron pair to produce any hadronic
final state, normalized to the QED  cross-section for production of a
muon-antimuon pair
\begin{equation}\label{eq:rdef}
R=\frac{\sigma(e^+e^-\to {\rm
    hadrons})}{\sigma(e^+e^-\to\mu^+\mu-)}.
\end{equation}
The numerator receives contributions from any Feynman diagram with
partons in the final state, see Fig.~\ref{fig:r}.

Because $R$ is a ratio of total cross-sections, it is dimensionless, and it depends only on the center-of-mass energy squared $s$
of the $e^+e^-$ collision. Hence, at energies
high enough that quark masses can be neglected, by dimensional
analysis, it can be only a function of the coupling with dimensionless
coefficients. In a classical theory one would conclude that these
coefficients must be numerical constants, but in the quantized theory
all predictions depend on a renormalization scale $\mu_R$, hence one
can form a dimensionless ratio  $s/\mu^2_R$ on which the coefficients
may also depend, so
$R=R\left(\frac{s}{\mu_R^2},\as(\mu_R)\right)$. However physical
predictions cannot depend on the choice of $\mu_R$ and thus $R$ must
satisfy the renormalization group equation
\begin {align}\label{eq:rge}
\mu_R^2\frac{d}{d\mu_R^2} R\left(\frac{s}{\mu_R^2},\as(\mu_R^2)\right)=\left[\mu_R^2\frac{\partial}{\partial\mu_R^2}+\beta(\as)\frac{\partial}{\partial\as}\right] R\left(\frac{s}{\mu_R^2},\as(\mu_R^2)\right)=0,
\end{align}
where we have introduced the  beta function
\begin{equation}\label{eq:betadef}
  \mu_R^2 \frac{d}{d \mu_R^2}
  \as (\mu_R^2)= \beta(\as (\mu_R^2))=-
  \beta_0\as^2+O(\as^3),
\end{equation}
which is a universal property of the theory, and perturbatively
computable.

Equation~(\ref{eq:rge}) is easy to solve directly, however the
general solution can be  found
based on a simple observation. Namely  that, because
$R$ does not depend on $\mu_R^2$, one can choose  $\mu^2_R=s$. With this
choice we find that
\begin{equation}\label{eq:rgsol}
  R=R\left(1,\as(s)\right),
  \end{equation}
namely that $R$ is a function of $\as(s)$ with numerical
coefficients. Integrating Eq.~(\ref{eq:betadef}) one finds that
$\as(s)$ in terms of 
$\as(\mu_R^2)$ is given by
\begin{equation}\label{eq:loas}
\as(s)=\frac{\as(\mu_R^2)}{1+\beta_0
  \as(\mu_R^2)\ln\frac{s}{\mu_R^2}},
\end{equation}
so with the choice $\mu^2_R=s$ we are exploiting renormalization group
invariance to sum  to all orders in $\as(\mu_R^2)$ logs of the scale
ratio $\frac{s}{\mu_F^2}$, which is large when $s$ is large.

Having worked out this simple case, we can now turn to the
exponentiation of collinear singularities. The starting point is the
factorized expression of the physical cross-section in terms of parton
distributions and a partonic cross-section. We consider  again for
definiteness  as an explicit example the Drell-Yan
process of Sect.~\ref{sec:massing}, but now with incoming protons with momenta
$P_1$ and $P_2$ and center of mass energy $s=(P_1+P_2)^2$. The
physically measurable cross section
depends only
on the invariant mass $Q^2$ of the final-state gauge boson and on the
dimensionless variable $\tau$ Eq.~(\ref{eq:taudef}).
Assuming that the two
incoming partons carry respectively fractions $x_1$, $x_2$ of the
momenta of the two incoming  protons the center-of-mass energy of the
partonic collision is $x_1x_2 s$, so by  momentum
conservation $x_1x_2 s=Q^2$. Integrating over the incoming parton
momentum fractions, the measurable cross-section $\sigma$ can thus be written as
\begin{equation}\label{eq:fact1}
\sigma(\tau)=\int_0^1 dz\int_0^1dx_1 \int_0^1 dx_2\, \delta(\tau- x_1 x_2 z)
q_1(x_1) q_2(x_2) \hat\sigma(z),
\end{equation}
where $q_i(x_i)$ denote the two PDFs, $\hat\sigma(z)$ is the
cross-section computed with incoming partons with momenta $p_i$
and with $z=\frac{Q^2}{2p_1\cdot p_2}$ (partonic cross-section, henceforth), and we have for the
time being not indicated explicitly the dependence of the various
quantities on $Q^2$ and the renormalization and factorization
scales. Also, for simplicity we have only indicated a single partonic
initial state, while in general the right-hand side of
Eq.~(\ref{eq:fact1}) will contain a sum over all possible
initial-state parton pairs that can lead to the desired final state.

The integral over the momentum fractions in Eq.~(\ref{eq:fact1}) is
actually a double convolution:
\begin{equation}\label{eq:fact2}
\sigma(\tau)=\int_\tau^1  \frac{dx_1}{x_1}\int_{\tau/x_1}^1 \frac{dx_2}{x_2 }q_1(x_1)
q_2(x_2) \hat\sigma\left(\frac{\tau}{x_1x_2}\right)=\int_\tau^1 \frac{dy}{y}
{\cal L} (y) \hat\sigma\left(\frac{\tau}{y}\right)= [{\cal L}\otimes
  \hat\sigma](\tau),
\end{equation}
with $\cal L$, the parton luminosity, defined as 
\begin{equation}\label{eq:lumi}
  {\cal L} (y)\equiv\int_y^1\frac{dx_1}{x_1}
q_1(y)q_2\left(\frac{y}{x_1}\right)= [q_1\otimes q_2](y).
\end{equation}
It follows that upon Mellin transformation the cross-section can be
written as the ordinary product of the  partonic cross-section and the
parton distributions.
Writing now explicitly the dependence on all
kinematic variables we have 
\begin{align}\label{eq:fact3}
\sigma(N,Q^2)&=\hat\sigma_0(Q^2)
C\left(\frac{Q^2}{\mu^2_F},N,\as(\mu^2_R),\frac{\mu^2_R}{\mu^2_F}\right)q_1(N,\mu^2_F)q_2(N,\mu^2_F)\\\label{eq:fact4}
&=\hat\sigma_0(Q^2)
C\left(\frac{Q^2}{\mu^2_F},N,\as(\mu^2_R),\frac{\mu^2_R}{\mu^2_F}\right) {\cal L} (N,\mu^2_F),
\end{align}
where
\begin{equation}\label{eq:lumin}
  {\cal L} (N,\mu^2_F)=q_1(N,\mu^2_F)q_2(N,\mu^2_F),
\end{equation}
and we have written the partonic
cross-section $\hat\sigma$ in terms of a coefficient function $C$ by factoring out its leading-order expression
$\hat\sigma_0(Q^2)$, which does not depend on the strong coupling as
the leading-order process is electroweak. The coefficient function
$C$ is consequently dimensionless and thus can only depend on dimensionless
ratios of dimensionful variables, that include $Q^2$, and the
renormalization and factorization scales that were respectively
introduced in order to define the renormalized theory and to treat the
collinear singularities.

In Eq.~(\ref{eq:fact3}) by slight abuse of notation we have used the
same symbol to denote a function and its Mellin transform:
\begin{align}\label{eq:melsig}
\sigma(N)=\int_0^1dx\, x^{N-1} \sigma(x);\\ 
\label{eq:melsighat}
\hat \sigma_0C(N)=\int_0^1dx\, x^{N-1} \hat\sigma(x);\\ 
\label{eq:melq}
q_{1,2}(N)=\int_0^1dx\, x^{N-1} q_{1,2}(x).
\end{align}
The physical cross-section can only depend on the physical
kinematic variables, i.e. $Q^2$, and the Mellin variable $N$, conjugate
to  $\tau$  Eq.~(\ref{eq:taudef}).  We have introduced a factorization scale according to the
argument of Sect.~\ref{sec:massing}, on which both the partonic
cross-section and the parton distributions depend, though the physical
cross-section does not. We have finally introduced a renormalization
scale $\mu_R$ according to the argument of the previous section, on
which the perturbatively computable coefficient function must depend
in order to compensate the dependence on it of the renormalized strong
coupling $\as(\mu_R^2)$ in such a way that the physical
cross-section does not depend on it.

We now impose the condition that physical observables must be
independent of the factorization scale:
\begin{equation}\label{eq:mufindep}
\mu_F^2\frac{\partial}{\partial\mu_F^2}\ln
\left[C\left(\frac{Q^2}{\mu^2_F},N,\as(\mu^2_R),\frac{\mu^2_R}{\mu^2_F}\right)
  {\cal L} (N,\mu^2_F)\right]=0.
\end{equation}
  The condition is imposed on the logarithmic derivative because we
  know from Sect.~\ref{sec:massing} that the dependence on the
  factorization scale is logarithmic, and we have imposed it on the
  logarithm of the physical cross-section because this immediately leads a 
  condition relating the coefficient function and the luminosity
  (u.e. the PDFs):
\begin{equation}\label{eq:mufindepsig}
\mu_F^2\frac{\partial}{\partial\mu_F^2}\ln
C\left(\frac{Q^2}{\mu^2_F},N,\as(\mu^2_R),\frac{\mu^2_R}{\mu^2_F}\right)=-\mu_F^2\frac{\partial}{\partial\mu_F^2}\ln
  {\cal L} (N,\mu^2_F)\equiv -2\gamma_{qq}(N,\as(\mu^2_R)).
\end{equation}
Furthermore, exploiting the independence of result on $\mu_R$,
we perform the partial derivative  with respect to $\mu_F^2$ in
Eq.~(\ref{eq:mufindepsig}) at
fixed $\frac{\mu_R}{\mu_F}$, because, with
          this choice, solving the renormalization group equation
          will simultaneously sum all large scale ratios.

In the last step we have  defined $\gamma_{qq}$ as the
logarithmic derivative of the quark PDF, and  the factor of 2 is a
consequence of the fact that the Mellin transform of the luminosity,
Eq.~\eqref{eq:lumi}, is the product of two quark PDFs.
%which must be equal but with the opposite sign. 
We have further
made use of the fact that $\gamma_{qq}$ cannot depend on $Q^2$: hence, on
dimensional grounds, it can only be a function of the strong coupling,
which, in turn, is evaluated at scale $\mu_R$ as a consequence of the
renormalization process. 
Comparing Eq.~(\ref{eq:mufindepsig}) to the collinear subtracted partonic
cross-section given by Eq.~(\ref{eq:sigmahat}), we identify $\gamma_{qq}$ with
the Mellin transformed expression~(\ref{eq:andim}) of the splitting
function.

We can thus view Eq.~(\ref{eq:mufindep}) as an equation satisfied by
the coefficient function: the Callan-Symanzik equation 
\begin{equation}\label{eq:cseq}
\mu^2_F\frac{\partial}{\partial\mu^2_F}
  C\left(\frac{Q^2}{\mu^2_F},N,\as(\mu^2_R),\frac{\mu^2_R}{\mu^2_F}\right)=-
  2\gamma_{qq}(N,\as(\mu_R^2))
  C\left(\frac{Q^2}{\mu^2_F},N,\as(\mu^2_R),\frac{\mu^2_R}{\mu^2_F}\right),
\end{equation}
with the partial derivative performed at fixed  $\frac{\mu_R}{\mu_F}$.

\subsection{Collinear resummation}
\label{sec:rgres}

The exponentiation of collinear logarithms follows from solving the
renormalization group equation, Eq.~(\ref{eq:cseq}). 
We can solve it in two steps. First, we trade the derivative with
respect to the factorization scale with a derivative with respect to
the physical scale, noting that the coefficient function for fixed
$\frac{\mu_R}{\mu_F}$ only depends on $\mu_F$ through the ratio
$\frac{Q^2}{\mu_F^2}$:
\begin{equation}\label{eq:csdm}
Q^2\frac{\partial}{\partial Q^2}
  C\left(\frac{Q^2}{\mu^2_F},N,\as(\mu^2_R),\frac{\mu^2_R}{\mu^2_F}\right)=  2\gamma_{qq}(N,\as(\mu_R^2))
  C\left(\frac{Q^2}{\mu^2_F},N,\as(\mu^2_R),\frac{\mu^2_R}{\mu^2_F}\right).
\end{equation}
Next, as we did when solving the renormalization group equation
satisfied by the $R$ ratio, we  exploit the independence of results
on the renormalization scale and choose $\mu_R^2=Q^2$. With this
choice we get
\begin{equation}\label{eq:rge2}
Q^2\frac{\partial}{\partial Q^2}
  C\left(\frac{Q^2}{\mu^2_F},N,\as(Q^2),\frac{Q^2}{\mu^2_F}\right)=
  2\gamma_{qq}(N,\as(Q^2))
  C\left(\frac{Q^2}{\mu^2_F},N,\as(Q^2),\frac{Q^2}{\mu^2_F}\right).
\end{equation}
This is now straightforward to solve, as it is a first order ordinary
differential equation, with solution
\begin{equation}\label{eq:cssol}
C \left(\frac{Q^2}{\mu^2_F},N,\as(Q^2),\frac{Q^2}{\mu^2_F}\right)=C\left(1,N,\as(Q^2),1\right)\exp \int_{\mu^2_F}^{Q^2}\frac{d\mu^2}{\mu^2}
2\gamma_{qq}(N,\as(\mu^2)).
\end{equation}

This solution manifestly displays the independence of physical
observables on the factorization scale: indeed, the dependence of $C
\left(\frac{Q^2}{\mu^2_F},N,\as(Q^2),\frac{Q^2}{\mu^2_F}\right)$
Eq.~(\ref{eq:cssol}) on the scale $\mu_F$ is through the lower extreme
of integration in the exponential, and this exactly matches the scale
dependence of the luminosity, as given by
Eq.~(\ref{eq:mufindepsig}). Note also that the further dependence on
the scale through $\alpha_s$, which is present in
Eq.~(\ref{eq:mufindepsig}) because the derivative is taken at fixed
$\frac{\mu_R}{\mu_F}$, has been reabsorbed when choosing the
renormalization scale $\mu_R^2=Q^2$, so that the derivative in
Eq.~(\ref{eq:rge2}) is a partial derivative, i.e. only acts on the
direct dependence of  $C
\left(\frac{Q^2}{\mu^2_F},N,\as(Q^2),\frac{Q^2}{\mu^2_F}\right)$ on
$\frac{Q^2}{\mu^2_F}$, for fixed $\alpha_s$. This generalizes the
procedure that we followed when solving the renormalization group
equation~(\ref{eq:rge}) which we can now view as the special case of the
Callan-Symanzik equation~(\ref{eq:cseq}) in which the anomalous
dimension vanishes.

This proves the exponentiation of the collinear logarithms: the
coefficient function is given by a function that depends on scale
only through the coupling (like the $R$ ratio), times the exponential
of an integral over scale that sums to all order collinear logarithms,
along with the running with the coupling. 

It is useful to rewrite the solution~(\ref{eq:cssol}) in an
equivalent form by changing variable of integration from $\mu_F^2$ to
$\as(\mu_F^2)$ using Eq.~(\ref{eq:betadef}):
\begin{equation}\label{eq:cssol1} 
\ln C\left(\frac{Q^2}{\mu_F^2},N,\as(Q^2),\frac{Q^2}{\mu^2_F}\right)= \ln C\left(1,N,\as(Q^2),1\right)+2\int_{\as(\mu_F^2)}^{\as(Q^2)}\frac{d\alpha}{\beta(\alpha)}
\gamma_{qq}(N,\alpha),
\end{equation}
where we have also written the logarithm of the coefficient function,
in order to ease the comparison with resummed results that we will
derive in the next section.
Using the leading-order beta function Eq.~(\ref{eq:betadef}) the
exponentiated collinear factor becomes
\begin{align}\label{eq:expcoll}
\exp\left[-\frac{2\gamma^{(0)}_N}{\beta_0}\int_{\as(\mu_F^2)}^{\as(Q^2)}\frac{d\alpha}{\alpha}\right]
&=\left(\frac{\as(Q^2)}{\as(\mu_F^2)}\right)^{-\frac{2\gamma_N^{(0)}}{\beta_0}}.
\end{align}
Further, using Eq.~(\ref{eq:loas}) to express $as(Q^2)$ in terms of
$\as(\mu_F^2)$
we get
\begin{align}
\label{eq:colll}
\exp\left[-\frac{2\gamma^{(0)}_N}{\beta_0}\int_{\as(\mu_F^2)}^{\as(Q^2)}\frac{d\alpha}{\alpha}\right]
 &=\left(1+\beta_0\as(\mu_F^2)\ln\frac{Q^2}{\mu_F^2}\right)^{\frac{2\gamma_N^{(0)}}{\beta_0}}+O\left[\alpha^2(\mu_F^2)\ln\frac{Q^2}{\mu_F^2}\right]\\\label{eq:colllo}
  &=1+2\as(\mu_F^2)\gamma_N^{(0)}\ln\frac{Q^2}{\mu_F^2}+O(\alpha^2(\mu_F^2)).
  \end{align}

Equation~(\ref{eq:colll}) shows explicitly that the expression
(\ref{eq:expcoll}) sums to all orders in $\as(\mu_F^2)$ the
collinear logs of $\frac{Q^2}{\mu_F^2}$: we have indeed exploited
renormalization group invaraince to sum all large scale ratios. On the
other hand,  Eq.~(\ref{eq:colllo})
shows that expanding this out to first order in $\as(\mu_F^2)$
leads back to the single-emission result Eq.~(\ref{eq:sigmahat}). 
Note that, because one can always express $\alpha_s(Q^2)$ in terms of
$\alpha_s(\mu_F^2)$ or conversely, we can equivalently view the
coefficient function as a function of $\frac{Q^2}{\mu_F^2}$ and either
$\alpha_s(\mu_F^2)$ or $\alpha_s(Q^2)$, as
Eqs.~(\ref{eq:cssol1}-\ref{eq:loas}) explicitly demonstrate.

\section{Soft resummation from RG invariance}
\label{sec:rgi}

Soft resummation, i.e. the all-order resummation and exponentiation of
the double soft-collinear logs discussed in Sect.~\ref{sec:sudakov},
follows from a similar argument to the collinear exponentiation
derived in the previous section. Namely, the real emission and virtual
contributions to the hard cross-section  discussed in
Sect.~\ref{sec:ircanc} can be argued to analogously factorize. The
infrared singularities then cancel between these two contributions,
which however depend on different kinematic variables: the real
emission depends on the maximum transverse momentum 
${k_t^2}_{\rm max}$, while the virtual contribution does not. Again,
dimensional analysis then implies that the cancellation of the
singularity must happen
through the scale dependence of the coupling, and it is calculable in
terms of it and of a universal function. Once integrated away, it
leads to all-order resummation and
exponentiation of logs of  ${k_t^2}_{\rm  max}$, which close to
threshold is a soft scale. We will discuss here the case of processes,
like deep-inelastic scattering, or the invariant mass distribution for
the Drell-Yan process, that only depend on a single dimensionful variable and
a dimensionless ratio. For the Drell-Yan process, that, following the
discussion in Part 1, we can take as a reference example, these
variables are 
the invariant
mass $Q^2$ of the final state gauge boson, and its  ratio $x=Q^2/s$ to the
center-of-mass energy $s$ of the collision

\subsection{The soft limit in Mellin space}
\label{sec:softlimm}
Because we wish to perform exponentiation  in Mellin space, where  the
phase space factorizes, we have first to understand how soft logs look
like in Mellin space.
To this purpose, we note that
\begin{align} \label{eq:melloga}
\int_0^1 dx\,x^{N-1} \left[\frac{\ln^p(1-x)}{1-x}\right]_+\,&=\frac{1}{(p+1)}\sum_{k=0}^{p+1}
\left(\begin{array}{c} p+1\\k \end{array}\right)\,\Gamma^{(k)}(1)\,
\left(\ln\frac{1}{N}\right)^{p+1-k}
+O\left(\frac{1}{N}\right)\\\label{eq:mellogb}
&=\frac{1}{p+1}\ln^{p+1}\frac{1}{N}+O\left(\ln^p \frac{1}{N}\right)+O\left(\frac{1}{N}\right) ,
\end{align}
where $\Gamma^{(k)}(1)$ denotes the $k$-th derivative of the Euler
Gamma function  $\Gamma(x)$ evaluated at $x=1$.
Equation~(\ref{eq:melloga}) is easy to prove by noting that
\begin{equation}\label{eq:melstir} 
\int_0^1 dx\,x^{N-1} (1-x)^{-1+\epsilon}=
\frac{\Gamma[\epsilon]\Gamma[N]}{\Gamma[N+\epsilon]}=
\frac{1}{\epsilon}\Gamma[1+\epsilon]N^{-\epsilon}\left[1+O\left(\frac{1}{N}\right)\right], 
\end{equation}
and expanding both sides about
$\epsilon=0$. Equation~(\ref{eq:melloga}) shows explicitly that, as
stated in Sect.~\ref{sec:ircanc}, a contribution of the form $\frac{1}{(1-x)_+}
$ is logarithmic upon integration with a test function: its Mellin
transform, up to power--suppressed corrections,  is just $\ln N$.

We further note that the $x\to1$ threshold limit corresponds to the
$\Re N\to\infty$ limit: indeed $dx x^{N-1}=d\left(\ln\frac{1}{x}\right)\exp\left(
-N\ln\frac{1}{x}\right)$ so in the $N\to\infty$ limit only the point at $x=1$
survives. It is in fact easy to prove that any real function has a
Mellin transform that vanishes as $N\to\infty$.
The partonic cross-section $\hat\sigma(x)$
however contains both contributions that are real functions of $x$,
but also contributions proportional to distributions localized at $x=1$,
like those discussed in Sect.~\ref{sec:ircanc}, whose Mellin transform
survive the $N\to\infty$ limit, as Eq.~(\ref{eq:melloga}) demonstrates. In fact
\begin{align} \label{eq:melconst}
  \int_0^1 dx\,x^{N-1} \delta(1-x)=1.
  \end{align}

We conclude that
the only contributions to the Mellin space partonic cross-section that do
not vanish as $N\to\infty$ are the Mellin transform of
distributions localized at $x=1$. The
Dirac delta distribution transforms into a constant, the
$\frac{1}{(1-x)_+}$ distribution leads to a single power of $\ln N$,
and each extra  power of $\ln(1-x)$  leads to  an extra power of $\ln
N$. Hence the Mellin transform of the highest power of $\ln(1-x)$ is
the same power of $\ln N$, with $\frac{1}{(1-x)_+}$ counting as the
first log. However, the Mellin transform also includes lower powers of
$\ln N$, all the way down to the constant.

The Mellin transform $\gamma_{qq}$ of the splitting function $P_{qq}$,
as given by Eq.~(\ref{eq:pqqfull}), is proportional to
$\frac{1}{(1-x)_+}$. It follows that for large $N$ it
behaves as $\ln N$ as $N\to\infty$, plus
a constant, up to terms that vanish as $N\to\infty$. Each collinear
log of $Q^2$ which is summed by the collinear exponentiation of
Eq.~(\ref{eq:colllo})  is consequently accompanied by $\ln N$. This is
an infrared log, corresponding to the $x\to1$ limit of the
splitting function. However, as discussed in Sect.~\ref{sec:ircanc},
the collinear log itself is 
\begin{equation}\label{eq:upperlog}
\ln \frac{{k_t^2}_{\rm max}}{\mu_F^2}=\ln \frac{Q^2}{\mu_F^2}\frac{(1-x)^2}{4 x}= \ln\frac{Q^2}{\mu_F^2}+
2\ln(1-x)+\hbox{non log},
\end{equation}
thereby producing an extra $\ln(1-x)$, and thus
leading, upon Mellin transformation, to an extra power of $\ln
N$. Hence each collinear emission leads to a power of $\ln^2N$ -- the
Sudakov double log, now in Mellin space. 

The goal of Sudakov resummation is thus to determine all
contributions to the partonic cross-section that survive the
$N\to\infty$ limit, specifically showing that they exponentiate. 

\subsection{Real-virtual factorization and the soft scale}
\label{sec:softlim}
Exponentiation of Sudakov double logs can be derived from a
resummation  argument  based on two ingredients. The first is the
factorization of the coefficient function into real emission and
virtual contribution, neither of which is separately finite in the
infrared, and the identification of the scales on which they depend. The
second is the renormalization group improvement of this scale
dependence.

\begin{figure}[!t]
  \centering
  \begin{minipage}{.29\linewidth}\begin{center}
      \includegraphics[width=\linewidth]{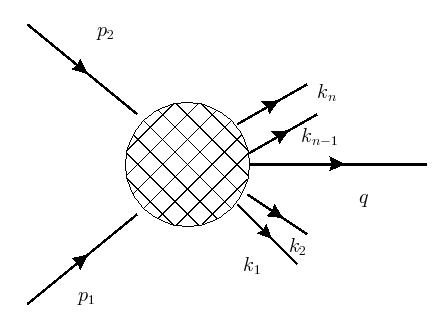}\end{center}\end{minipage}   \begin{minipage}{.19\linewidth}\begin{center}\vglue.15truecm
      {\Huge $=$ } \end{center}\end{minipage} 
\begin{minipage}{.29\linewidth}\begin{center}
\includegraphics[width=\linewidth]{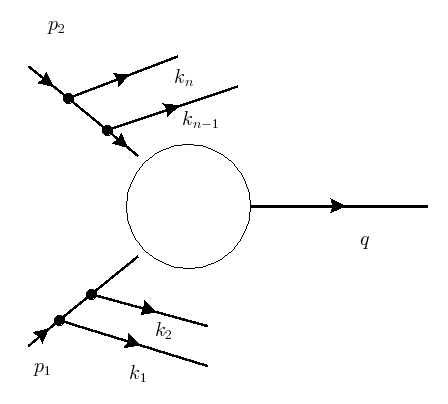}\end{center}\end{minipage} 
\begin{minipage}{.19\linewidth}\begin{center}\vglue.15truecm
      {\Large $+O(1-x)$ } \end{center}\end{minipage}
\caption{Graphical representation of the suppression of radiation from
  internal lines in the soft limit. It is assumed that the line
  with momentum $q$ is colorless and cannot radiate like in the
  Drell-Yan process.}    
 \label{fig:rvfact}
\end{figure}
The factorization is based on the (partly conjectural) fact that
real emission from the internal lines of a diagram is regular in the
infrared: i.e., in comparison to the eikonal contribution~(\ref{eq:softeik}), it does not diverge as $\frac{1}{p\cdot k}$ as
$k\to0$. In fact, we have already tacitly assumed this to be the case
when stating, towards the end of Sect.~\ref{sec:sudakov}, that the
argument presented there does not depend on the form of the amplitude
$M_0$. Indeed, for a generic amplitude $M_0$, on top of the contribution from radiation from external lines shown in
Fig.~\ref{fig:dyfull}, there are in general further
real-emission contributions coming from the internal lines of $M_0$,
that we neglected in the argument of Sect.~\ref{sec:sudakov}. The
suppression of radiation from internal lines then leads to
factorization, because soft radiation from external lines factorizes
due to the eikonal argument of Sect.~\ref{sec:soft}. 
This suppression  is relatively easy to prove in QED:
the argument, originally given in Ref.~\cite{Yennie:1961ad}, is summarized in Appendix~\ref{sec:yfs}. A general proof in QCD
is rather less straightforward, though the result is likely to hold at
least with processes with a colorless final state. The result is
represented pictorially for the Drell-Yan process in 
Fig.~\ref{fig:rvfact}.

Upon this assumption, the partonic cross-section factorizes as
\begin{equation}\label{eq:rvfac}
  \hat\sigma(Q^2,x;\epsilon)=H(Q^2;\epsilon)J(Q^2,x;\epsilon)[1+O(1-x)],
\end{equation}
where $H(Q^2;\epsilon)$ includes all the loop corrections, it has therefore
the kinematics of the leading-order process, and thus depends only on
$Q^2$, while $J(Q^2,x;\epsilon)$ is a dimensionless function that
includes the radiation from external
lines and thus the nontrivial kinematic dependence on $x=\frac{Q^2}{s}$, with $s$ the center-of-mass energy of the partonic
collision. We have
explicitly indicated a dependence on a (dimensional) regulator
$\epsilon$ because $H$ and $J$ are separately divergent, hence
Eq.~(\ref{eq:rvfac}) only makes sense at a regularized level.

Furthermore, recalling the argument from Sects.~\ref{sec:massing}-\ref{sec:ircanc},
we note that the radiation contribution
$J(Q^2,x;\epsilon)$, up to non-logarithmic terms, can only
depend on the kinematic variables $Q^2$ and $x$ through the upper
limit ${k_t^2}_{\rm  max}$ of the integration over the transverse
momentum of the emitted partons. This means that $J$ actually depends
on a fixed combination of the kinematic variables $Q^2$ and $x$, namely
by ${k_t^2}_{\rm  max}(Q^2,x)$.
The value of   ${k_t^2}_{\rm  max}$ for the Drell-Yan process,
already given in Eq.~(\ref{eq:ktmaxy}), can be determined as
follows. In the general case in which in the final state we have a
gauge boson with momentum $q$ and a set of partons with total momentum
$k'$, momentum conservation implies
\begin{equation}\label{eq:momcon}
  p_1+p_2=k'+q,
\end{equation}
where $p_i$ are the momenta of the incoming partons and $q^2=Q^2$.
In the center-of-mass reference frame the energy of the collision is given by
\begin{equation}\label{eq:cmass}
s=(p_1+p_2)^2=(q_0+k'_0)^2=\left(
\sqrt{Q^2+q_t^2+q_z^2}+\sqrt{{k'}^2+q_t^2+q_z^2}\right)^2,
\end{equation}
where in the last step $q_t=|\vec q_t|$ and $q_z$ are respectively the
transverse and longitudinal components of the momentum of the gauge
boson.  It follows that for fixed $s$, $q_t$ is maximum when
${k'}^2=q_z^2=0$. In this case,
\begin{equation}\label{eq:sqtmax}
s=\left(
     \sqrt{Q^2+{q_t^2}_{\rm max}}+{q_t^2}_{\rm max}\right)^2.
\end{equation}
Solving for ${q_t^2}_{\rm max}$, and noting that of course
${k_t^2}_{\rm  max}={q_t^2}_{\rm max}$,
we get
\begin{equation}\label{eq:qtmaxs}
  {k_t^2}_{\rm  max}={q_t^2}_{\rm max}=Q^2\frac{(1-x)^2}{4x},
\end{equation}
where $x=\frac{Q^2}{s}$, as anticipated in Eq.~(\ref{eq:ktmaxy}).

The resummation is performed in Mellin space, where, of
course, the coefficient function, given by Eq.~(\ref{eq:melsighat}), in turn
factorizes into the product of a function of $Q^2$ and a function that
depends on a fixed combination of $Q^2$ and
$N$.
Equation~(\ref{eq:melstir}) immediately implies that  with
${k_t^2}_{\rm  max}=Q^2\frac{(1-x)^2}{4x}$ the Mellin space coefficient function
only depends on $N$ through the soft 
scale
\begin{equation}\label{eq:softscale}\Lambda^2_{\rm soft}=
  \frac{Q^2}{N^2},
\end{equation}
up to corrections that vanish when $N\to\infty$. This in turn implies
that the Mellin transform of $J(Q^2,x;\epsilon)$, which contains all the $N$ dependence
of the coefficient function, only depends on $Q^2$ and $N$ through
$\Lambda^2_{\rm soft}$. Hence, upon Mellin transformation
$H(Q^2;\epsilon)$, which is 
$N$-independent, remains unchanged;  $J(Q^2,x;\epsilon)$ is
transformed into a function  of $ \frac{Q^2}{N^2}$; and the
Mellin-space coefficient
function is the product of these two factors.

\subsection{Renormalization group improving the soft scale}
\label{sec:physad}

The renormalization group argument that leads to resummation
is performed in terms of a physical anomalous dimension, defined as
\begin{equation}\label{eq:gamp}
 \gamma^{\rm phys} (N,\as(Q^2))\equiv Q^2\frac{d}{dQ^2}
\ln  C\left(\frac{Q^2}{\mu^2_F},N,\as(Q^2),\frac{Q^2}{\mu^2_F}\right).
  \end{equation}
Note that it follows from Eq.~(\ref{eq:cssol}) that $\gamma^{\rm phys}$
does not depend on the factorization scale $\mu_F$. This in particular
means that even though the coefficient function itself diverges when
$\mu_F\to0$, and indeed $\mu_F$ was introduced in
Sect.~\ref{sec:massing} in order to regulate the collinear
singularity, the scale dependence  of the log of the coefficient
function remains finite in the limit. This result is clear from the
expression of the coefficient function Eq.~(\ref{eq:cssol}), and it is
thus a direct consequence of the renormalization group
equation~(\ref{eq:mufindep}) that it satisfies.
Note also that  $\gamma^{\rm phys}$ differs from the anomalous dimension $2\gamma_{qq}$: it
provides the dependence of the physically measurable cross-section on
the physical scale $Q^2$ (hence its name), and it also receives a
contribution from the first factor on the right-hand side of
Eq.~(\ref{eq:cssol}).

Exploiting the factorization Eq.~(\ref{eq:rvfac}), the physical
anomalous dimension can be written
as a sum of two contributions,
which correspond respectively to the logarithmic derivative with
respect to the physical scale of the  virtual  and  real factors, $H$ and
$J$ respectively, with the latter computed in Mellin space (and the
former $N$-independent). Even though the physical anomalous dimension is finite, the real
and virtual contributions to it are not separately finite, and thus
this decomposition can only be performed in terms of regularized
quantities, which requires introducing a scale. It is the independence of the physical anomalous dimension
of this scale that leads to soft resummation, just like the
independence of the physical cross-section of the factorization scale
led to collinear resummation in Sect.~\ref{sec:rgres}.

Indeed, recalling the definition~(\ref{eq:melsighat}) of the coefficient
function, and substituting the Mellin-space version of the  factorized
expression of the partonic cross-section of Eq.~(\ref{eq:rvfac}) in the
definition~(\ref{eq:gamp}) of the physical anomalous dimension, we get
\begin{equation}\label{eq:pcl}
   \gamma^{\rm phys}(N,\as(Q^2))=\lim_{\epsilon\to0}\left[ \gamma^{(c)}\left(\frac{Q^2}{\mu^2},\as(\mu^2);\epsilon\right)
   +\gamma^{(l)}
     \left(\frac{Q^2/N^2}{\mu^2},\as(\mu^2);\epsilon\right)\right],
\end{equation}
where $\gamma^{(c)}=Q^2\frac{d}{dQ^2} \ln  H$ and
$\gamma^{(l)}=Q^2\frac{d}{dQ^2} \ln  J$, and all
expressions on the right-hand side are computed in $d$ dimensions, so
necessarily $\as$ must be $\mu$-dependent. Also, we have made use
of the fact that
$H$ and correspondingly $\gamma^{(c)}$ only depends on    $Q^2$, while
$J$, and correspondingly $\gamma^{(l)}$, only depends on
$\frac{Q^2}{N^2}$.
Because $\gamma^{(c)}$ and $\gamma^{(l)}$ are
both dimensionless, they can depend on $Q^2$ or respectively $Q^2/N$   only through
their ratio to  the regularization scale $\mu^2$. The $N$ dependence,
and thus the logs, is contained in  $\gamma^{(l)}$, while
$\gamma^{(c)}$ is constant, i.e. $N$--independent.

Following the argument of Sect.~\ref{sec:rgres} we could now introduce
a suitable scale, analogous to the factorization scale,  in order to renormalize $\gamma^{(l)}$ and
$\gamma^{(c)}$ in such a way that they become simultaneously finite in the
infrared. However, this is not necessary, as we are really only
interested in their sum, namely the physical anomalous
dimension, and we can instead
exploit the condition of $\mu$-independence of  $\gamma^{\rm phys}$:
\begin{equation}\label{eq:rgecl}
  \lim_{\epsilon\to0}\left[\mu^2\frac{d}{d\mu^2}
  \gamma^{(l)}\left(\frac{Q^2/N^2}{\mu^2},\as(\mu^2);\epsilon\right)+\mu^2\frac{d}{d\mu^2}
  \gamma^{(c)}\left(\frac{Q^2}{\mu^2},\as(\mu^2);\epsilon\right)\right]=0.
\end{equation}
But now we note that $\gamma^{(c)}$ and $\gamma^{(l)}$ depend on different
kinematic variables, hence the only way Eq.~(\ref{eq:rgecl}) can hold
is if in the $\epsilon\to0$ limit the scale derivatives  of both
$\gamma^{(c)}$ and $\gamma^{(l)}$ , that must be equal up to the sign, are
functions of $\as(\mu^2)$ with coefficients that do not depend on
either  $Q^2$ or $\frac{Q^2}{N^2}$:
\begin{equation}\label{eq:rgg}
  \lim_{\epsilon\to0}  \mu^2\frac{d}{d\mu^2}
  \gamma^{(l)}\left(\frac{Q^2/N^2}{\mu^2},\as(\mu^2);\epsilon\right)=-\lim_{\epsilon\to0}
  \mu^2\frac{d}{d\mu^2}
  \gamma^{(c)}\left(\frac{Q^2}{\mu^2},\as(\mu^2);\epsilon\right)=-g(\as(\mu^2)),
\end{equation}
with
\begin{equation}\label{eq:gexp}
  g(\as)= g_1 \as+g_2 \alpha^2_s+\dots.
  \end{equation}
We also note that the coefficients $g_i$ are necessarily finite,
as we shall see explicitly shortly.

We now observe that Eq.~(\ref{eq:rgg}) looks exactly like the
renormalization group equation~(\ref{eq:mufindepsig}) satisfied by the
logarithm of the coefficient function: it can be consequently solved
in the same way, leading to a solution of the form of
Eq.~(\ref{eq:cssol1}), both for $\gamma^{(l)}$ and $\gamma^{(c)}$, which only
differ because of the scale on which they depend -- and the sign of
their scale dependence. We get
\begin{align}\label{eq:gammacsol}
 \gamma^{(c)}\left(\frac{Q^2}{\mu^2},\as(\mu^2);\epsilon\right)&=
g_0^{(c)}\left(\as(Q^2);\epsilon\right)-\int_{\mu^2}^{Q^2}
\frac{d\lambda^2}{\lambda^2}g\left(\as(\lambda^2)\right)
\\
\label{eq:gammalsol}
\gamma^{(l)}\left(\frac{Q^2/N^2}{\mu^2},\as(\mu^2);\epsilon\right)&=
    g_0^{(l)}\left(\as(Q^2/N^2);\epsilon\right)+\int_{\mu^2}^{Q^2/N^2}
    \frac{d\lambda^2}{\lambda^2}g\left(\as(\lambda^2)\right).
    \end{align}
Substituting these solutions in the expression of Eq.~(\ref{eq:pcl}) of
the physical anomalous dimension we find  
\begin{equation}\label{eq:rgisol}
\gamma^{\rm phys}\left(N,\as(Q^2)\right)=
\lim_{\epsilon\to 0}\left[ g^{(c)}_0(\as(Q^2);\epsilon)+g^{(l)}_0\left(\as(Q^2/N^2);\epsilon\right)\right] +\int_{Q^2}^{Q^2/N^2} \frac{d\lambda^2}{\lambda^2} g(\as(\lambda^2)).
\end{equation}
This is manifestly independent of the scale $\mu$, and furthermore it
shows that because $\gamma^{\rm phys}$ is finite, both the function
$g(\as)$, and the sum of the initial conditions $g^{(c)}_0(\as(Q^2);\epsilon)$
and $g^{(l)}_0\left(\as(Q^2/N^2);\epsilon\right)$
must also be finite.

We can rewrite the solution Eq.~(\ref{eq:rgisol}) in a more compact
form by noting that we can relate the strong coupling at different
scales by integrating up the beta function Eq.~(\ref{eq:betadef}):  
\begin{align}\label{eq:finalres}
\gamma^{\rm phys}\left(N,\as(Q^2)\right)&=
\lim_{\epsilon\to0} \left[ g^{(c)}_0(\as(Q^2);\epsilon)+g^{(l)}_0\left(\as(Q^2);\epsilon\right)\right] 
  \nonumber\\
  &+\int_{Q^2}^{Q^2/N^2}
 \frac{d\lambda^2}{\lambda^2}\left[
   g(\as(\lambda^2))+\beta(\as(\lambda^2))\frac{dg^{(l)}_0}{d\as}(\as(\lambda^2))\right]\nonumber\\
 &=\bar g_0(\as(Q^2))+\int_{Q^2}^{Q^2/N^2}
 \frac{d\lambda^2}{\lambda^2} \bar g (\as(\lambda^2)).
\end{align}
where $\bar g_0$ and $\bar g$ are perturbative series in
$\alpha_s$.
This expression of the physical anomalous dimension is our final
result. Renormalization group invariance and the cancellation of
infrared singularities between factorized real emission contribution, that only
depends on the soft scale $\Lambda^2_{\rm soft}=\frac{Q^2}{N^2}$, and virtual contribution, that only depends
on the hard scale $Q^2$, leads to a prediction of the full logarithmic
dependence on the Mellin variable $N$ to all orders in $\as(Q^2)$.
As we shall see shortly, knowledge of the first-order coefficient in
the expansion of $\bar g$ fully determines leading log resummation,
knowledge of the next coefficient together with the first correction to
$\bar g_0$  determines the next-to-leading log result, and so on. These coefficients can
be determined by comparing to fixed-order results, thereby allowing
prediction of logarithmic terms to all orders based on fixed-order
knowledge.

\section{The resummed coefficient function}
\label{sec:rescf}
The resummed result given in  Eq.~(\ref{eq:finalres}) provides an expression for
the physical anomalous dimension, i.e. essentially for the scale
dependence of the coefficient function. Hence, a little more work is
required in order to arrive at a resummed expression for the
coefficient function itself. This is done by separating off the part
of the $Q^2$ dependence of the coefficient function that goes through
the ratio to the factorization scale $Q^2/\mu_F^2$. The final resummed
result can then be cast in various equivalent forms, and shown to sum
to all orders in $\alpha_s(Q^2)$ contributions corresponding to a
given fixed logarithmic
accuracy by including a suitable finite number of terms in the
expansion of the coefficients that enter its expression.

\subsection{The soft scale and the factorization scale}
\label{sec:avsb}

In order to go from the physical anomalous dimension to the coefficient
function we start recalling
that contributions to the coefficient function
that survive  in the large-$N$ limit are either constant, i.e. $N$-independent functions of
$\alpha_s$ only, or logarithmic, i.e. powers of $\ln N$. The
two contributions to the physical anomalous dimension on the right-hand side of
Eq.~(\ref{eq:finalres}) then respectively provide us with an
expression for the scale dependence of constant and logarithmic
contributions to the coefficient function.
Namely, we can write the coefficient function as
\begin{equation}\label{eq:cffact}
    C\left(N,\frac{Q^2}{\mu^2_F},\as(Q^2)\right)=C^{(c)}\left(\as(Q^2)\right)C^{(l)}\left(\frac{Q^2/N^2}{\mu^2_F},\as(Q^2)\right)+O\left(\frac{1}{N}\right)
  \end{equation}
  where
  \begin{align}
\label{eq:pancc}
Q^2\frac{d}{dQ^2}C^{(c)}\left(\as(Q^2)\right)&=\bar g_0(\as(Q^2)) C^{(c)}\left(\as(Q^2)\right)\\
\label{eq:pancl}
   Q^2\frac{d}{d Q^2}  C^{(l)}\left(\frac{Q^2/N^2}{\mu^2_F},\as(Q^2)\right)&=
   \left[\int_{Q^2}^{Q^2/N^2}
 \frac{d\lambda^2}{\lambda^2} \bar g (\as(\lambda^2))\right]C^{(l)}\left(\frac{Q^2/N^2}{\mu^2_F},\as(Q^2)\right).
  \end{align}
  The two factors $C^{(c)}$ and $C^{(l)}$ in the right-hand side of
  Eq.~(\ref{eq:cffact}) can be thought of as a providing a
  factorization of the coefficient function analogous to that of Eq.~(\ref{eq:rvfac}) that we
  started from, with the function of
  the soft scale $C^{(l)}$ playing the role of $J$ and the function of the hard scale
  $C^{(c)}$ playing the role of $H$. However, the two factors are now individually
  finite.

Specifically, Eq.~(\ref{eq:pancl}) tells us that logarithmically enhanced terms, contained in $C^{(l)}$, satisfy
\begin{align}
\ln C^{(l)}\left(\frac{Q^2/N^2}{\mu^2_F},\as(Q^2)\right)-\ln
C^{(l)}\left(\frac{Q_0^2/N^2}{\mu^2_F},\as(Q_0^2)\right)&=
\int_{Q_0^2}^{Q^2}\frac{dk^2}{k^2}\int_{k^2}^{k^2/N^2}
 \frac{d\lambda^2}{\lambda^2} \bar g (\as(\lambda^2))
\nonumber\\
&= -\int_1^{N^2} \frac{dn}{n}\, 
\int_{Q_0^2/n}^{Q^2/n}\frac{dk^2}{k^2}\,\bar{g}\left(\as(k^2)\right),
\label{eq:clscal}
\end{align}
where $Q_0^2$ is some reference value of the hard scale $Q^2$. We may be
tempted to interpret $Q_0$ as the factorization scale $\mu_F$, but then
we realize that, in actual fact, the coefficient function has the form
given by Eq.~(\ref{eq:cssol}): hence the logarithmic part of the
coefficient function is in turn given by the product of two factors,
one of which depends on the factorization scale, and the other which
does not.

In order to cast the result Eq.~(\ref{eq:clscal}) in the desired form,
we let 
\begin{equation}\label{eq:abdec}
\bar g(\as)=A(\as)-\frac{\partial B(\as(k^2))}{\partial\ln k^2}
\end{equation}
so that, substituting in Eq.~(\ref{eq:clscal}), $C^{(l)}$
takes the form
\begin{equation}\label{eq:ab1}
C^{(l)}\left(\frac{Q^2/N^2}{\mu^2_F},\as(Q^2)\right)=\exp\left\{\int_{1}^{N^2}
 \frac{dn}{n}\left[\left(-\int_{\mu^2(\mu_F)}^{Q^2/n}\frac{dk^2}{k^2}
A(\as(k^2))\right)+ B(\as(Q^2/n))\right]\right\},
\end{equation}
where $\mu^2(\mu_F)$ is some reference scale that depends on $\mu_F$.

We can determine both the value of the scale $\mu$ and of the function
$A(\as(k^2))$ in Eq.~(\ref{eq:ab1}) by demanding that the $\mu_F$
dependence of $C^{(l)}$ be given by the renormalization group
equation~(\ref{eq:cseq}), as we now show. First, we note that
Eq.~(\ref{eq:pqqfull}), together with
the Mellin transformation formula~(\ref{eq:melloga}), implies that
$\gamma_{qq}$ is proportional to $\ln N$, up to constants and terms
that vanish as $N\to\infty$. This turns out to be the case to all
perturbative orders (in the $\overline{\rm MS}$ factorization scheme):
\begin{equation}\label{eq:cusp}
\gamma_{qq}(N,\as(\mu^2))=-\ln N\sum_k \left(\frac{\as(\mu^2)}{\pi}\right)^k
A_k+O\(N^0\)+O\(\frac{1}{N}\).
\end{equation}
The coefficient of the logarithmically enhanced contribution to the
anomalous dimension is known as cusp anomalous dimension. This
logarithmically enhanced contribution is present in the $\gamma_{qq}$
and $\gamma_{gg}$ anomalous dimensions, that respectively characterize
collinear emission of a gluon from a quark or a gluon line. Collinear emission of a
quark from a gluon or a quark is not logarithmically enhanced, and in
fact the corresponding anomalous dimensions vanish as $N\to\infty$.  For the
Drell-Yan process, on which we are focusing, only the quark cusp anomalous
dimension is thus relevant.

The desired result now follows noting that if we let $\mu^2(\mu_F)=\mu_F^2$ in Eq.~(\ref{eq:ab1}) we
immediately get
\begin{equation}\label{eq:diffcl}
\mu_F^2\frac{d}{d\mu_F^2} \ln
C^{(l)}\(\frac{Q^2/N^2}{\mu_F^2},\as(Q^2)\)=\ln N^2 A(\as(\mu_F^2)).
\end{equation}
Note the factor of 2 from $\ln N^2=2\ln N$ exactly provides the factor of 2 needed to obtain twice $-\gamma_{qq}$ 
as in Eq.~(\ref{eq:mufindepsig}).
Comparing to Eq.~(\ref{eq:cseq}) with the anomalous
dimension given by Eq.~(\ref{eq:cusp}) (since  $C^{(l)}$ only includes
logarithmically enhanced contributions) we find
that the resummation function $A(\as(\mu^2))$ coincides with the cusp
anomalous dimension:
\begin{equation}\label{eq:cuspad}
A(\as)=\sum_k \left(\frac{\as}{\pi}\right)^k
A_k.
\end{equation}

\subsection{The resummed result and its accuracy}
\label{sec:equiv}

We are now ready to present the resummed coefficient function and
study the accuracy that the resummation has achieved.
The final expression for the
resummed coefficient function, collecting all results, is
\begin{equation}\label{eq:finrescf}
C \left(N,\frac{Q^2}{\mu_F^2},\as(Q^2)\right)=C^{(c)}\left(\as(Q^2)\right)\exp\int_{1}^{N^2}
 \frac{dn}{n}\left[\left(-\int_{\mu_F^2}^{Q^2/n}\frac{dk^2}{k^2}
A(\as(k^2))\right)+ B(\as(Q^2/n))\right],
\end{equation}
with $A$ given by the cusp anomalous dimension of Eq.~(\ref{eq:cuspad}),
and the functions $C^{(c)}$ and $B$ given as power series in
$\alpha_s$:
\begin{align}\label{eq:clexp}
 C^{(c)}\left(\as\right) &= \left(\frac{\as}{\pi}\right)^k C^{(c)}_k\\
\label{eq:bexp}
B\left(\as\right) &=\sum_k   \left(\frac{\as}{\pi}\right)^k B_k.
\end{align}

This resummed result shows that the double logs associated to the multiple emissions
of Sect.~\ref{sec:order} do indeed exponentiate: the cusp anomalous
dimension is the most singular part of the splitting function in the
infrared limit, and the splitting function is in turn the coefficient
of the most singular contribution in the collinear limit, with the
infrared and collinear singularities respectively regulated by the
plus prescription and the factorization scale $\mu_F$. Indeed,
\begin{equation}\label{eq:aone}
  A_1=C_F,
\end{equation}
the coefficient of the double log Eq.~(\ref{eq:suddl}), that now is exponentiated.
Higher-order contributions to the cusp anomalous dimension correspond to
double logs associated to higher-order corrections to the splitting
function and the anomalous dimension  in this soft-collinear
limit. Finally, the function $B(\alpha_s)$ resums to all orders
contributions that are associated to soft radiation, but not in the
ordered collinear region of Sect.~\ref{sec:order}.

The leading logs, which correspond to an extra power of log squared at
each order in $\alpha_s$ come from the $A_1$ contribution to
Eq.~(\ref{eq:finrescf}), while $B$ starts contributing at the
next-to-leading logarithmic order. Also, beyond leading log one must
include contributions to the function $C^{(l)}$, because even though
not logarithmically enhanced, by interference with logarithmically
enhanced contributions coming from $A$ and $B$ they produce
subleading logarithmic contributions.

An efficient way of keeping track of the logarithmic orders is to
write the coefficient function as
\begin{equation}\label{eq:cfgs}
C(N,\as)=g_0(\as) 
\exp\left[\frac{1}{\alpha_s} g_1(\as\ln N)+g_2(\as\ln N)+ \as g_3(\as\ln
  N)+\dots\right],
\end{equation}
where all functions $g_i$ are power series in their respective
arguments:
\begin{equation}\label{eq:gexpdef}
  g_i(x)=\sum_{k=k_i^{\rm min}}^\infty g^k_{i} x^k.
\end{equation}
Here $k_0^{\rm min}=0$, $k_1^{\rm min}=2$ and $k_i^{\rm min}=1$ for all
$i\ge 2$: the coefficient function starts with $1$
at order $\alpha_s^0$ (as it must be, given that the leading order
$\sigma_0$ has been factored out, Eq.~(\ref{eq:melsighat})), and the
$g_1$ term starts at $O(\alpha_s)$ with a double log.

\begin{table}[!t]
  \centering
  \begin{tabular}[c]{c c c c c c c}
  \midrule
{ log accuracy} & accuracy of $C$: &  accuracy of $\ln C$: &$g_0$: &$g_j$
 &$A$:&$B$:\\
 &   $\as^nL^k$ &  $\as^nL^k$  &  $\as^k$ & order &  $\as^i$  &  $\as^i$ \\
\midrule
    LL & $k=2n$ & $k= n+1$ & 0 &1 & 1& 0\\
    NLL & $2n-2\leq k\leq 2n$ & $k=n$ & 1& 2& 2& 1\\ 
    NNLL & $2n-4\leq k\leq 2n$& $k=n-1$ & 2& 3 & 3 &2 \\
\midrule
\end{tabular}
\caption{Summary of the coefficients in the resummation formulae
  required to achieve a given logarithmic accuracy.}
 \label{tab:pertord}
\end{table}
Subsequent contributions $g_i$ in the exponent correspond to the leading,
next-to-leading,\dots, logarithmic approximation, where at the leading
order the power of $\alpha_s$ is always by unit lower  than the power
of $\ln N$, at next-to-leading it is the same, at
next-to-next-to-leading one more and so on. In the coefficient
function at the leading logarithmic level each order in $\alpha_s$ is
accompanied by an extra power of $\ln^2N$, and it is fully predicted
by knowledge of $g_1$, which in turn is entirely predicted by
knowledge of the $A_1$ coefficient, i.e. the leading-order cusp
anomalous dimension. The function $g_2$ is predicted by knowledge of
$A_2$ and $B_1$. Knowledge of $g_2$ then predicts at each order in $\alpha_s$
the coefficient of the next-lower order power of $\ln N$. However, if this is supplemented
also by the first order coefficient $g^1_0$ in the expansion of the
prefactor function $g_0$,  at each order in $\alpha_s$ the
coefficients of two
next lower powers of $\ln N$ are actually predicted. This is usually referred to as
next-to-leading logarithmic (NLL) approximation in the QCD literature.
Confusingly in the SCET literature it is usually called NLL',
while NLL is the approximation without the $g^1_0$ coefficient, in
which the coefficient of one less power of $\ln N$ is predicted. The
pattern continues at next orders, and it is summarized in Table~\ref{tab:pertord}.

The coefficients $A_k$, $B_k$ and  $C^{(c)}_k$ can be determined by
matching the resummed result to a fixed-order
computation. Specifically, the leading cusp anomalous dimension $A_1$
is determined by the coefficient of the highest power of log of a
next-to-leading (NLO) order computation. Comparison to the NLO also
determines the coefficients $B_1$ and   $C^{(c)}_1$ that enter the
next-to-leading logarithmic (NLL) resummation. However, the NLO cusp
anomalous dimension $A_2$ only appears in a NNLO fixed order
computation. It is easy to check that this pattern persists to all
orders: the coefficients  $B_k$ and  $C^{(c)}_k$ that enter the
N$^k$LL resummation are fully determined by a fixed N$^k$LO fixed
order result, but the  N$^k$LO cusp anomalous dimension only appears
in a  N$^{k+1}$LO fixed order computation. On the other hand, the cusp
anomalous dimension is process-independent, so once the cusp anomalous
dimension is known to  N$^k$LO, a N$^k$LO fixed order computation of a
process fully determines the coefficients needed for N$^k$LL resummation.

The resummed result of Eq.~(\ref{eq:finrescf}) takes an especially simple
form when choosing a factorization scale that coincides with the hard
scale, $\mu_F^2=Q^2$. With this choice the exponential evolution
factor is absent in the expression~(\ref{eq:cssol}) of the
coefficient function, because it is entirely reabsorbed in the parton
luminosity, so the coefficient function becomes a function of $N$ and
$\as(Q^2)$ only. In this case its resummed expression becomes
\begin{align}\label{eq:finrescfa}
C (N,1,\as(Q^2))&=C^{(c)}\left(\as(Q^2)\right)\exp\int_{1}^{N^2}
 \frac{dn}{n}\left[\left(-\int_{Q^2}^{Q^2/n}\frac{dk^2}{k^2}
A(\as(k^2))\right)
+ B(\as(Q^2/n))\right].
\end{align}
This can be rewritten in an interesting way by switching the order of
the two integrations and performing the $n$ integral:
\begin{align}\label{eq:reslog}
C (N,1,\as(Q^2))&=C^{(c)}\left(\as(Q^2)\right)\exp\int_{Q^2}^{Q^2/N^2}\frac{dk^2}{k^2}
\left[A(\as(k^2))\int_{N^2}^{Q^2/k^2}\frac{dn}{n}+ B(\as(k^2))\right]
\\
&=C^{(c)}\left(\as(Q^2)\right)
   \exp\int_{Q^2}^{Q^2/N^2}
 \frac{dk^2}{k^2}\left[
A(\as(k^2))\ln\frac{Q^2/N^2}{k^2}
+  B(\as(k^2))\right].
\label{eq:reslog2}
\end{align}
This form of the resummed result shows that the $A$ term effectively resums logs by performing a further
perturbative evolution with the large $N$ anomalous dimension
given by Eq.~(\ref{eq:cusp}) from the hard scale $Q^2$ to the soft scale $Q^2/N^2$.

This can be made even more explicit by rewriting the resummation
exponent using the identity
\begin{equation}
\int_0^1dx\,\frac{x^{N-1}-1}{1-x}\ln^k(1-x)
=-\sum_{n=0}^\infty\frac{\Gamma^{(n)}(1)}{n!}\frac{d^n}{dL^n}\int_0^{1-1/N}\frac{dx}{1-x}\ln^k(1-x)+O(1/N)
\label{eq:melid}
\end{equation}
where
\begin{equation}
L=\ln\frac{1}{N}.
\end{equation}
Using this in Eq.~(\ref{eq:reslog}) we get
\begin{align}\label{eq:catanires}
  C (N,1,\as(Q^2))=&C^{(c)}\left(\as(Q^2)\right)\nonumber\\\times&\exp\left[\int_0^1 \! dx\,
  \frac{x^{N-1}-1}{1-x}\left(\int_{Q^2}^{\left(1-x\right)^2 Q^2}
  \frac{dk^2}{k^2} 2A(\as(k^2)+\bar B(\as((1-x)^2 Q^2))\right)\right],
\end{align}
where the function $\bar  B$ is a series in $\alpha_s$ whose
coefficients are determined order by order using Eq.~(\ref{eq:melid})
in terms of the coefficients $A_k$ and $B_k$: for instance $\bar
B_1=B_1+2\gamma_E A_1$ and so on.
This is the form of the resummation that was given in the original
paper~\cite{Catani:1989ne}. The term proportional to $A$ in the
exponent is just the Mellin transform of $P_{qq}(\alpha_s(k^2))$,
integrated from scale $Q^2$ to scale $Q^2/N$, thereby exposing the
physical meaning of the resummation. On the other hand, it should be
noted that the integral over $k^2$ in Eq.~(\ref{eq:catanires}) is
ill-defined, because the strong coupling $\alpha_s(k^2)$
diverges at the so-called Landau pole, i.e. when
$k^2$ is small enough that the denominator in Eq.~(\ref{eq:loas})
vanishes. Hence, Eq.~(\ref{eq:catanires}) is only meaningful insofar
as it is equal to Eq.~(\ref{eq:reslog}) up to terms that vanish in
the $N\to\infty$ limit.

\section{Transverse momentum resummation}
\label{sec:ptres}

The resummation formalism that we have discussed so far is
conceptually interesting and transparent, but of limited
phenomenological interest, because measured cross-sections vanish at
threshold, and become very small close to threshold. Hence in the
region where resummation effects are significant the measured
cross-section is quite small. A related resummation formalism which is
instead very relevant phenomenologically is that for
transverse-momentum distributions, such as the Drell-Yan
cross-section, but now differential not only with respect to the gauge
boson mass, but also with respect to its transverse momentum $q_t$. This can be
thought of as the differential counterpart of the cross-section that
we discussed so far. The transverse momentum distribution if
computed at fixed perturbative order diverges in the $q_t\to0$
limit. This divergence is akin to the divergence of  the real emission
cross-section of Sect.~\ref{sec:sudakov}: it is related to a
soft-collinear Sudakov double-log, that can be similarly resummed to all orders.

A treatment of transverse momentum resummation goes beyond the scope of these
lectures. It will suffice here to make some general remarks and quote
its final form. The same sort of argument that led us to perform
threshold resummation in Mellin space
leads to transverse momentum resummation  in Fourier space, conjugate to
transverse momentum  itself. Indeed, the real emission contribution to
the transverse momentum distribution of the gauge boson respects
the conservation of transverse momentum and consequently the phase
space $d\Phi_n$ for $n$ gluon emission contains a momentum conservation delta:
\begin{equation}\label{eq:ktcons}
  d\Phi_n\propto d^2k_{t1}d^2k_{t2}\dots d^2k_{tn}\delta^{(2)}\left(\vec k_{t1}+\dots+\vec k_{tn}-\vec q_t\right),
\end{equation}
where  $k_t^i$ are the transverse momenta of the emitted gluons and
$q_t$ is the transverse momentum of the gauge boson.

Resummation requires as a first step the factorization of the real
emission contributions which are then exponentiated, but factorization
of the phase space is broken by the transverse momentum conserving
delta. Factorization of the phase space is achieved  by performing a Fourier
transformation, just like (recall Sect.~\ref{sec:phsp}) the
factorization of longitudinal phase space (i.e. the integrals over the
momentum fractions $x_i$ Eq.~(\ref{eq:conv})) is achieved through
Mellin transformation. Indeed, we have
\begin{align}\label{eq:trphspa}
d^2 q_td^2k_{t1}\dots d^2k_{tn}\delta(\vec k_{t1}+\dots+\vec k_{tn}-\vec q_t)
&= d^2q_t \int\frac{d^2b}{(2\pi)^2}e^{i\vec b\cdot\vec q_t}  d^2k_{t1}e^{-i\vec b\cdot\vec k_{t1}}\dots d^2k_{tn}e^{-i\vec b\cdot\vec k_t^n}\\
    \label{eq:trphspb}
&=d^2q_t\int\frac{d|\vec b|^2}{4\pi}J_0(|\vec b| |\vec q_t|)  d^2 k_{t1}e^{-i\vec b\cdot\vec k_{t1}}\dots d^2k_{tn}e^{-i\vec b\cdot\vec k_t^n},
\end{align}
where in the last step we have performed the integration over the
azimuthal angle of $\vec q_t$ and $J_0$ is a Bessel function. Because
transverse momentum is two-dimensional and longitudinal momentum
one-dimensional, transverse momentum resummation is
accordingly somewhat more cumbersome than threshold resummation.

A second complication is related to the fact that in transverse
momentum resummation the soft and collinear logs arise as logarithms
of $q_t^2$ which, as we saw in Sect.~\ref{sec:coll}, are present for
all kinds of parton emission, hence all partonic subchannels
contribute, and not only the diagonal channel as in the case of
threshold resummation. With these preliminary considerations, we are
now ready to present the expression for the resummed transverse
momentum distribution in the $ij$ partonic subchannel, with
factorization scale set equal to $Q^2$:
\begin{align}\label{eq:ptres}
\frac{d\hat{\sigma}_{ij}}{dq_t^2}\left(N,q_t,\as\left(Q^2\right),Q^2\right)
&=\sigma_0 
\int_0^{\infty} db\,\frac{b}{2}\,J_0\left(b q_t\right) 
H_{ij}\left(N,\as\left(Q^2\right)\right)\nonumber\\
&\quad \exp\left[-\int_{\frac{1}{b^2}}^{Q^2} \frac{dq^2}{q^2}\left[
    A^{q_t}\left(\as\left(q^2\right)\right)\ln\frac{Q^2}{q^2}+{ B^{q_t}}
    \left(\as\left(q^2\right),N\right)\right]\right].
\end{align}
Note that the cross-section still depends on $Q^2$ and the
dimensionless ratio $x=Q^2/s$, and now also on $q_t$, and the
resummation is performed in Fourier-Mellin space in order to factorize
both the longitudinal and the transverse phase space, so the resummed
result depends on $N$ conjugate to $x$ and $b$ conjugate to $q_t$.

The similarity of Eq.~(\ref{eq:ptres}) to the threshold resummation of  Eq.~(\ref{eq:reslog})
should be clear. The soft resummation scale is now $1/b^2$, and the
exponent takes the form of an integral from the hard scale $Q^2$ to
this soft scale, with a double logarithmic contribution $A^{q_t}$ and a
single-logarithmic contribution $B^{q_t}$. Note that $A^{q_t}$
coincides with the cusp anomalous dimension only up to order
$\alpha^3$, but beyond this order (corresponding to NNLL resummation)
it differs from it.
The functions  $H_{ij}\left(N,\as(M^2)\right)$ do not depend on
the soft resummation scale $1/b^2$, and 
are thus analogous to the prefactor $C^{(c)}\left(\as(Q^2)\right)$ in
the threshold resummation formula. They are universal functions which
do not depend on the specific process and only depend on the parton
flavors $ij$.

\section{Conclusion}
\label{sec:con}
In these lectures we have provided a simple introduction to basic,
elementary concepts of soft resummation. The lectures barely scratch the
surface of what has become, since the seminal papers of more than
thirty years ago, a vast and active research subject. A subject  to which, among
other things,  several workshops are devoted, and that has close 
connection to the Monte Carlo parton showers, that are
fundamental for collider phenomenology, and that implement resummation
numerically.

On top of the transverse momentum resummation that we have briefly
touched upon in Sect.~\ref{sec:ptres}, other applications of the
resummation formalism, actively pursued in recent years,
are related to the resummation of observables
with hadronic final states and thus depend on color, specifically those that
characterize jets and their
substructure~\cite{Luisoni:2015xha,Marzani:2019hun}.  
Also, as
mentioned in the introduction, in the past two decades many resummation
results have been re-derived, and some derived for the first time,
using the formalism of soft-collinear effective theory, in which the
fundamental QCD Lagrangian is replaced by an effective Lagrangian in
which the fundamental fields are decomposed as the sum of fields with
different scaling properties that describe  hard, soft and collinear
excitations~\cite{Becher:2014oda}.

We hope that this brief introduction will stimulate the readers to
deepen their knowledge of this fascinating, old but still very active research  topic.

\bigskip
{\bf Acknowledgements:} S.F. thanks Micha\l\ Prasza\l owicz for
inviting him to lecture at the Zakopane school, all the students who
attended the lectures for their interest and questions, and
G.~Korchemsky and E.~Ruiz Arriola for interesting discussions during
the school. This work has received funding from the European Union
NextGeneration EU program - NRP Mission 4 Component 2 Investment 1.1 -
MUR PRIN 2022 - CUP G53D23001100006 through the Italian Ministry of
University and Research (MUR).

The Feynman diagrams included in these lectures have been produced
using the {\tt FeynGame} package~\cite{Harlander:2020cyh,Harlander:2024qbn,Bundgen:2025utt}.

\appendix

\section{Emission from internal lines}
\label{sec:yfs}

We provide a brief summary of the classic argument from
Ref.~\cite{Yennie:1961ad} that proves that in QED only radiation from
external lines leads to infrared divergences.

We consider a generic amplitude in QED with an incoming
fermion line carrying momentum $p$, an outgoing fermion line carrying
momentum $p'$, and a number of external photons with momenta $q_i$:
\begin{equation}
M_0=\bar u(p')\Gamma(p,q)u(p).
\end{equation}
Some of the photons with momenta $q_1,\ldots,q_n$ may be virtual; in such case, an integral over the loop momentum is
understood.  We now consider the same amplitude, with one additional
emitted photon with momentum $k$ in the soft $k\to 0$ limit.  The additional photon may be emitted either by the incoming
fermion line, or by the outgoing fermion line, or by some internal
line. Thus
\begin{equation}
M_1=M_1^\mu\epsilon^*_\mu(k),
\end{equation}
with
\begin{equation}
M_1^\mu=
\bar u(p')\left[
\gamma^\mu\frac{\slashed p'+\slashed k+m}{(p'+k)^2-m^2}\Gamma(p,q)
+\Gamma(p-k,q)\frac{\slashed p-\slashed k+m}{(p-k)^2-m^2}\gamma^\mu
+\Gamma^\mu(p,q,k)\right] u(p).
\label{m1def}
\end{equation}
Some slight redefinition of the momenta $q_i$ is needed in order
to keep $p$ and $p'$ at the same values, while emitting one more
photon, but this is irrelevant in the small-$k$ limit. 
We now observe that the QED Ward identity
\begin{equation}
k_\mu M_1^\mu=0
\end{equation}
gives
\begin{equation}
\Gamma(p-k,q)=\Gamma(p,q)+k_\mu\Gamma^\mu(p,q,k).
\label{wardid}
\end{equation}
Equation~\eqref{wardid} can be used to
eliminate $\Gamma(p-k,q)$ from Eq.~\eqref{m1def}.  We obtain
\begin{align}
M_1^\mu&=
\bar u(p')\Bigg[
\gamma^\mu\frac{\slashed p'+\slashed k+m}{(p'+k)^2-m^2}\Gamma(p,q)
+\Gamma(p,q)\frac{\slashed p-\slashed k+m}{(p-k)^2-m^2}\gamma^\mu
\nonumber\\
&+\Gamma_\nu(p,q,k) \left(g^{\nu\mu}+k^\nu\frac{\slashed p-\slashed k+m}{(p-k)^2-m^2}\gamma^\mu
\right)\Bigg]u(p).
\label{m1def1}
\end{align}
We may now adopt the procedure that led to Eq.~(\ref{eq:softeik}), to get
\begin{equation}
M_1^\mu=
\bar u(p')\left[
\Gamma(p,q)\left(\frac{{p'}^\mu}{p'\cdot k}-\frac{{p}^\mu}{p\cdot k}\right)
+\Gamma_\nu(p,q,k) \left(g^{\nu\mu}-\frac{k^\nu p^\mu}{p\cdot k}
\right)\right] u(p)\left[1+O(k)\right].
\label{m1def2}
\end{equation}
The first term in Eq.~(\ref{m1def2}) is the usual eikonal approximation for the emission of
one additional photon in the soft limit.  We now want to show that the
second term is suppressed in comparison to the eikonal term when
$k\to 0$. It is not a priori obvious that this is  the case.  Indeed,
assume that the
photon with momentum $k$ is emitted by an internal fermion line
carrying momentum $p+Q$, where $Q$ is some partial sum of the $q_i$.
We may write
\begin{equation}
\Gamma(p,q)=\Gamma_2(p,q)\frac{1}{\slashed p +\slashed Q -m}\Gamma_1(p,q)
\label{gamma}
\end{equation}
and
\begin{equation} 
\Gamma_\nu(p,q,k)
=\Gamma_2(p-k,q)\frac{1}{\slashed p +\slashed Q -\ks-m}\gamma_\nu\frac{1}{\slashed p +\slashed Q -m}
\Gamma_1(p,q).
\label{gammamu}
\end{equation} 
If both $Q$ and $k$ are small, $\Gamma_\nu$ contains two small denominators,
compared to the single one in Eq.~(\ref{gamma}); as a consequence,
the second term in Eq.~(\ref{m1def2}) may be as singular as the first one.

We now show that this is not the case, thanks to a cancellation
between the two terms in round brackets in the second term
of Eq.~(\ref{m1def2}). We have
\begin{align}
&
\frac{1}{\slashed p +\slashed Q -\ks-m}\gamma_\nu\left(g^{\mu\nu}-\frac{p^\mu k^\nu}{p\cdot k}\right)
=
\frac{\slashed p +\slashed Q -\ks+m}{(p+Q-k)^2-m^2}
\gamma_\nu\left(g^{\mu\nu}-\frac{p^\mu k^\nu}{p\cdot k}\right)
\nonumber\\
&\qquad=\frac{1}{(p+Q-k)^2-m^2}
\nonumber\\
&\qquad\quad
\left[-\gamma_\nu(\slashed p +\slashed Q -\slashed k-m)
+2(p+Q-k)_\nu\right]\left(g^{\mu\nu}-\frac{p^\mu k^\nu}{p\cdot k}\right)
%\nonumber\\
%&\qquad=\frac{1}{(p+Q-k)^2-m^2}
%\nonumber\\
%&\qquad\quad
%\left[-\left(\gamma^\mu-\frac{p^\mu\ks}{p\cdot k}\right)(\slashed p +\slashed Q -m)
%-\ks\gamma^\mu+2(p+Q)^\mu
%-2\frac{p^\mu}{p\cdot k}(p\cdot k+Q\cdot k)\right]
\nonumber\\
&\qquad=\frac{1}{(p+Q-k)^2-m^2}
\nonumber\\
&\qquad\quad
\left[-\gamma_\nu\left(g^{\mu\nu}-\frac{p^\mu k^\nu}{p\cdot k}\right)(\slashed p +\slashed Q -m)
-\ks\gamma^\mu+2Q^\mu-2p^\mu\frac{Q\cdot k}{p\cdot k}\right].
\label{eq:YFS}
\end{align}
The factor of $\slashed p +\slashed Q -m$ in the first term cancels one of
the potentially small denominators in Eq.~(\ref{gammamu});
the remaining terms are proportional to either $k$ or $Q$, and therefore
soften the overall singularity. This argument
holds irrespective of the fermion mass $m$, and therefore in particular
when $m=0$, which is the case of interest in QCD. 

We conclude that the last term in Eq.~(\ref{m1def2}) is at most as singular
as $M_0$, and therefore indeed less singular than the eikonal term.
A crucial role is played by the Ward identity:
in the last step in Eq.~\eqref{eq:YFS} a cancellation between two factors of $p^\mu$
has taken place. 
Quite obviously, a direct extension of the same argument to QCD is not
possible. Gluons are emitted both from fermion and gluon lines;
furthermore, the structure of Slavnov-Taylor identities is far more
complicated in the non-abelian case.

\bibliographystyle{UTPstyle}
\bibliography{zakores}

@article{Harlander:2020cyh,
    author = "Harlander, R. V. and Klein, S. Y. and Lipp, M.",
    title = "{FeynGame}",
    eprint = "2003.00896",
    archivePrefix = "arXiv",
    primaryClass = "physics.ed-ph",
    reportNumber = "TTK-20-04",
    doi = "10.1016/j.cpc.2020.107465",
    journal = "Comput. Phys. Commun.",
    volume = "256",
    pages = "107465",
    year = "2020"
}

@article{Harlander:2024qbn,
    author = "Harlander, Robert and Klein, Sven Yannick and Schaaf, Magnus C.",
    title = "{FeynGame-2.1 -- Feynman diagrams made easy}",
    eprint = "2401.12778",
    archivePrefix = "arXiv",
    primaryClass = "hep-ph",
    reportNumber = "TTK-24-05",
    doi = "10.22323/1.449.0657",
    journal = "PoS",
    volume = "EPS-HEP2023",
    pages = "657",
    year = "2024"
}

@article{Bundgen:2025utt,
    author = {B{\"u}ndgen, Lars and Harlander, Robert V. and Klein, Sven Yannick and Schaaf, Magnus C.},
    title = "{FeynGame 3.0}",
    eprint = "2501.04651",
    archivePrefix = "arXiv",
    primaryClass = "hep-ph",
    reportNumber = "TTK-24-56, P3H-24-096",
    doi = "10.1016/j.cpc.2025.109662",
    journal = "Comput. Phys. Commun.",
    volume = "314",
    pages = "109662",
    year = "2025"
}

@article{Yennie:1961ad,
    author = "Yennie, D. R. and Frautschi, Steven C. and Suura, H.",
    title = "{The infrared divergence phenomena and high-energy processes}",
    doi = "10.1016/0003-4916(61)90151-8",
    journal = "Annals Phys.",
    volume = "13",
    pages = "379--452",
    year = "1961"
}

@book{Sterman:1993hfp,
    author = "Sterman, George F.",
    title = "{An Introduction to quantum field theory}",
    isbn = "978-0-521-31132-8",
    publisher = "Cambridge University Press",
    month = "8",
    year = "1993"
}

@article{White:2015wha,
    author = "White, C. D.",
    title = "{An Introduction to Webs}",
    eprint = "1507.02167",
    archivePrefix = "arXiv",
    primaryClass = "hep-ph",
    doi = "10.1088/0954-3899/43/3/033002",
    journal = "J. Phys. G",
    volume = "43",
    number = "3",
    pages = "033002",
    year = "2016"
}

@book{Becher:2014oda,
    author = "Becher, Thomas and Broggio, Alessandro and Ferroglia, Andrea",
    title = "{Introduction to Soft-Collinear Effective Theory}",
    eprint = "1410.1892",
    archivePrefix = "arXiv",
    primaryClass = "hep-ph",
    reportNumber = "PSI-PR-14-12",
    doi = "10.1007/978-3-319-14848-9",
    publisher = "Springer",
    volume = "896",
    year = "2015"
}

@book{Peskin:1995ev,
    author = "Peskin, Michael E. and Schroeder, Daniel V.",
    title = "{An Introduction to quantum field theory}",
    doi = "10.1201/9780429503559",
    isbn = "978-0-201-50397-5, 978-0-429-50355-9, 978-0-429-49417-8",
    publisher = "Addison-Wesley",
    address = "Reading, USA",
    year = "1995"
}

@book{Collins:2011zzd,
    author = "Collins, John",
    title = "{Foundations of Perturbative QCD}",
    doi = "10.1017/9781009401845",
    isbn = "978-1-009-40184-5, 978-1-009-40183-8, 978-1-009-40182-1",
    publisher = "Cambridge University Press",
    volume = "32",
    year = "2011"
}

@book{Berestetskii:1982qgu,
    author = "Berestetskii, V. B. and Lifshitz, E. M. and Pitaevskii, L. P.",
    title = "{Quantum Electrodynamics}",
    isbn = "978-0-7506-3371-0",
    publisher = "Pergamon Press",
    address = "Oxford",
    series = "Course of Theoretical Physics",
    volume = "4",
    year = "1982"
}

@article{Sterman:1986aj,
    author = "Sterman, George F.",
    title = "{Summation of Large Corrections to Short Distance Hadronic Cross-Sections}",
    reportNumber = "Print-86-0873 (IAS,PRINCETON)",
    doi = "10.1016/0550-3213(87)90258-6",
    journal = "Nucl. Phys. B",
    volume = "281",
    pages = "310--364",
    year = "1987"
}

@article{Contopanagos:1996nh,
    author = "Contopanagos, Harry and Laenen, Eric and Sterman, George F.",
    title = "{Sudakov factorization and resummation}",
    eprint = "hep-ph/9604313",
    archivePrefix = "arXiv",
    reportNumber = "ANL-HEP-25, ANL-HEP-PR-96-25, CERN-TH-96-75, ITP-SB-96-17",
    doi = "10.1016/S0550-3213(96)00567-6",
    journal = "Nucl. Phys. B",
    volume = "484",
    pages = "303--330",
    year = "1997"
}

@article{Duhr:2025cye,
    author = {Duhr, Claude and Gardi, Einan and Jaskiewicz, Sebastian and L{\"u}bken, Jonas and Vernazza, Leonardo},
    title = "{Infrared singularities and the collinear limits of multi-leg scattering amplitudes}",
    eprint = "2507.21854",
    archivePrefix = "arXiv",
    primaryClass = "hep-ph",
    reportNumber = "BONN-TH-2025-25",
    month = "7",
    year = "2025"
}

@article{Forte:2002ni,
    author = "Forte, Stefano and Ridolfi, Giovanni",
    title = "{Renormalization group approach to soft gluon resummation}",
    eprint = "hep-ph/0209154",
    archivePrefix = "arXiv",
    reportNumber = "RM3-TH-02-02, GEF-TH-11-02",
    doi = "10.1016/S0550-3213(02)01034-9",
    journal = "Nucl. Phys. B",
    volume = "650",
    pages = "229--270",
    year = "2003"
}

@book{Weinberg:1995mt,
    author = "Weinberg, Steven",
    title = "{The Quantum theory of fields. Vol. 1: Foundations}",
    doi = "10.1017/CBO9781139644167",
    isbn = "978-0-521-67053-1, 978-0-511-25204-4",
    publisher = "Cambridge University Press",
    month = "6",
    year = "2005"
}

@article{Luisoni:2015xha,
    author = "Luisoni, Gionata and Marzani, Simone",
    title = "{QCD resummation for hadronic final states}",
    eprint = "1505.04084",
    archivePrefix = "arXiv",
    primaryClass = "hep-ph",
    reportNumber = "MIT-CTP-4672, MPP-2015-102",
    doi = "10.1088/0954-3899/42/10/103101",
    journal = "J. Phys. G",
    volume = "42",
    number = "10",
    pages = "103101",
    year = "2015"
}

@book{Marzani:2019hun,
    author = "Marzani, Simone and Soyez, Gregory and Spannowsky, Michael",
    title = "{Looking inside jets: an introduction to jet substructure and boosted-object phenomenology}",
    eprint = "1901.10342",
    archivePrefix = "arXiv",
    primaryClass = "hep-ph",
    doi = "10.1007/978-3-030-15709-8",
    publisher = "Springer",
    volume = "958",
    year = "2019"
}

@article{Catani:1989ne,
    author = "Catani, S. and Trentadue, L.",
    title = "{Resummation of the QCD Perturbative Series for Hard Processes}",
    reportNumber = "DFF-93/3/89",
    doi = "10.1016/0550-3213(89)90273-3",
    journal = "Nucl. Phys. B",
    volume = "327",
    pages = "323--352",
    year = "1989"
}

@book{Ellis:1996mzs,
    author = "Ellis, R. Keith and Stirling, W. James and Webber, B. R.",
    title = "{QCD and collider physics}",
    doi = "10.1017/CBO9780511628788",
    isbn = "978-0-511-82328-2, 978-0-521-54589-1",
    publisher = "Cambridge University Press",
    volume = "8",
    month = "2",
    year = "2011"
}

\end{document}